\documentclass[useAMS,usenatbib]{mn2e}
\usepackage{journals}
\usepackage[pdftex]{graphicx,color}
\usepackage[latin1]{inputenc}
\usepackage[T1]{fontenc}
\usepackage{graphics}
\usepackage{amsfonts}
\usepackage{amsmath}
\usepackage{multicol}
\usepackage{layout}
\usepackage{amssymb}
\usepackage{natbib}
\usepackage[english]{babel}

\title[Weak lensing BigMultiDark Simulation database release]
      {MultiDarkLens Simulations: weak lensing light-cones and data base presentation}
      \author[Giocoli et al. 2015] 
             {\parbox{\textwidth}{Carlo Giocoli$^{1}$\thanks{E-mail:
                   \href{mailto:carlo.giocoli@lam.fr}
                        {carlo.giocoli@lam.fr}}, 
                 Eric Jullo$^1$,
                 R. Benton Metcalf$^2$,  
                 Sylvain de la Torre$^1$,
                 Gustavo Yepes$^3$,
                 Francisco Prada$^{4,5,6}$ 
                 Johan Comparat$^{3,4}$,             
                 Stefan G\"ottlober$^7$,                
                 Anatoly Kyplin$^8$,
                 Jean-Paul Kneib$^{9,1}$,               
                 Margarita Petkova$^{10}$,                
                 HuanYuan Shan$^9$,
                 Nicolas Tessore$^2$ \\ }\\
               $^1$ Aix Marseille  Universit\'e,  CNRS,  LAM (Laboratoire  d'Astrophysique  de
Marseille)  UMR 7326,  13388, Marseille,  France \\
               $^2$ Dipartimento di Fisica e Astronomia, Alma Mater Studiorum Universit\`{a} di 
               Bologna, viale Berti Pichat, 6/2, 40127 Bologna, Italy\\ 
               $^3$ Departamento de F\'{\i}sica Te\'orica, Universidad Aut\'onoma de Madrid, Cantoblanco, 28049, Madrid, Spain \\ 
               $^4$ Instituto de F\'{\i}sica Te\'orica, (UAM/CSIC), Universidad Aut\'onoma de Madrid,  Cantoblanco, E-28049 Madrid, Spain \\
               $^5$ Campus of International Excellence UAM+CSIC, Cantoblanco, E-28049 Madrid, Spain \\
               $^6$ Instituto de Astrof\'{\i}sica de Andaluc\'{\i}a (CSIC), Glorieta de la Astronom\'{\i}a, E-18080 Granada, Spain \\
               $^7$ Leibniz-Institut fur Astrophysik (AIP), An der Sternwarte 16, D-14482 Potsdam, Germany \\
               $^8$ Astronomy Department, New Mexico State University, MSC 4500, PO Box 30001, Las Cruces, NM, 880003-8001, USA\\
               $^9$ Laboratoire d'astrophysique (LASTRO), Ecole Polytechnique  F\'ed\'erale de Lausanne (EPFL), \\ 
               \hspace{0.2cm} de Sauverny, CH-1290 Versoix, Switzerland \\
               $^{10}$ Department of Physics, Ludwig-Maximilians-Universitaet, Scheinerstr. 1, D-81679 Muenchen, Germany\\
             }
\begin{document}
\date{}
\maketitle
\label{firstpage}
\pagerange{\pageref{firstpage}--\pageref{lastpage}} \pubyear{2015}
\begin{abstract}
  
In this paper we present a  large database of weak lensing light cones
constructed  using   different  snapshots   from  the   Big  MultiDark
simulation (BigMDPL). The ray-tracing through  different multiple plane has been
performed  with the  \textsc{GLAMER}  code accounting  both  for single  source
redshifts  and for  sources distributed  along the  cosmic time.  This
first paper presents  weak lensing forecasts and  results according to
the geometry of the VIPERS-W1 and VIPERS-W4 field of view.  Additional
fields will be available on our database  and new ones can be run upon
request.  Our database also contains  some tools for lensing analysis.
In this  paper we present  results for convergence power  spectra, one
point and high order weak  lensing statistics useful for forecasts and
for cosmological studies. Covariance  matrices have also been computed
for the different realisations of the W1 and W4 fields. In addition we
compute  also   galaxy-shear  and  projected  density   contrasts  for
different halo  masses at  two lens redshift  according to  the CFHTLS
source redshift distribution both using stacking and cross-correlation
techniques, finding very good agreement.

\end{abstract}
\begin{keywords}
  galaxies:  halos  -  cosmology:  theory  - dark  matter  -  methods:
  analytical - gravitational lensing: weak
\end{keywords}

\section{Introduction}

Weak  gravitational lensing  is fast  becoming an  important tool  for
measuring  the evolution  in the  expansions of  the Universe  and the
distribution of  matter within it.   Large scale imaging  surveys that
are currently  being carried out such  as DES \citep{des05,flaugher05}
and surveys that  are planned for the future such  as LSST, Euclid and
WFIRST \citep{euclidredbook,wfirst,ivezic08} will  use weak lensing to
test theories of dark energy,  dark matter and alternatives to General
Relativity  with unprecedented  accuracy.  As  these new  weak lensing
measurements become more precise,  systematic effects arising from the
measurement  of   gravitational  shear  from  galaxy   images  and  in
translating  these measurements  into  constraints  on cosmology  will
become more and more important.

The simplest way to confront  theory with weak lensing observations is
through a second  order statistic such as the shear  power spectrum or
correlation function as  a function of scale and  source redshift.  In
the linear  regime of  structure formation  and under  the assumptions
that the  lensing is weak  and the  Born approximation is  valid these
statistics  can be  predicted straightforwardly,  as will  be reviewed
later  (for a  review of  weak lensing  see \cite{bartelmann01}).   As
lensing observations become more precise  the predictions need to take
into account  more complicated effects.   In practice the  redshift of
individual  galaxies   need  to  be  measured   photometrically  which
introduces errors and outliers and  the surveyed area is a complicated
shape with  many masked regions which  introduces correlations between
the measurements of  the shear power spectrum  or correlation function
at different scales which need to be measured precisely.  In addition,
the gravitational evolution of structure on the relevant scales is not
entirely linear,  but includes nonlinear structures  and even baryonic
effects.  Then the  lensing itself on a galaxy-by-galaxy  bases is not
always weak  in the  sense that  higher order  lensing effects  can be
safely ignored.  These effects can  only be addressed with simulations
although  in  some  cases  analytic  methods  can  be  used  that  are
calibrated or fit  to simulations.  The difficulties  in measuring the
shear from individual galaxy images is another important, but separate
problem that is not directly related to what is discussed here.

In this  paper we present  the first set of  a series of  weak lensing
simulations  that will  be provided  to  the community  through a  web
portal.  The intention is to  create high quality simulations that can
be used  by any group to  improve techniques and analyse  survey data.
The underlying cosmological simulation is the BigMDPL simulation
\citep{prada14}.  The light-cone  construction and ray-tracing through
this     simulation    is     done     with     the    \textsc{GLAMER}     code
\citep{metcalf14,petkova14}.   Shear  maps,  shear catalogs  and  some
analysis  tools are  provided.  The  BigMDPL is  big enough  to
provide many independent light-cones and its resolution is high enough
to resolve  the relevant nonlinear structures.   \textsc{GLAMER} calculates the
light  paths, shear  and  convergence without  resorting  to the  weak
lensing   approximation  or   the  Born   (unperturbed  light   paths)
approximation  so  that  the  impact of  such  approximations  can  be
evaluated.  The first fields reported on  here are in the shape of the
W1 and W4 fields observed in the VIPERS survey \citep{guzzo14}.

The  paper is  organised as  follows: in  section \ref{sectionsim}  we
describe the cosmological simulations used; in section \ref{secmethod}
we present  the methodology adopted  in constructing the  lens density
maps  as  well  as  our   multi-plane  ray-tracing  code.  In  section
\ref{sectionstat}  we  present  our  weak  lensing  results.   Section
\ref{sumandcon} is a summary of conclusions and discussion.

\section{The Big MultiDark Simulation}
\label{sectionsim} 

In this work  we perform full ray-tracing simulations  with the matter
density  distributions extracted  from  the  
BigMDPL\footnote{\url{http://www.multidark.org}, \url{https://www.cosmosim.org/cms/simulations/multidark-project/bigmdpl}}  simulation
\citep{prada14}.   This  simulation has  been  performed  to meet  the
science requirements  of the  BOSS galaxy  survey, i.e.  the numerical
requirements  for mass  and force  resolution that  allow one  to well
resolve those haloes and subhaloes  that can host typical BOSS massive
galaxies  at  $z\sim 0.5$.   This  allows  for  the creation  of  mock
catalogs with  appropriate galaxy bias and  clustering. The simulation
is in a $\mathrm{\Lambda}$CDM universe comprised of $3840^3$ particles
in a  box of 2.5 comoving  Gpc/$h$ on a side.  Initial conditions have
been      generated     at      redshift     $z_{init}=100$      using
\textsc{GINNUNGAGAP}\footnote{\url{https://github.com/ginnungagapgroup/ginnungagap}}
publicly available  full MPI-OpenMP initial conditions  generator code
that  uses  Zeldovich  approximation   with  an number  of
particles only limited by the computer resources. The simulation  has been run with the  L-GADGET-2 code (see
\citet{klypin14}, for details). The  cosmological parameters have been
chosen  to be  consistent  with the  latest fits  to  the Planck  data
\citep{planckxvi}.    The    mass    and   force    resolutions    are
$2.36 ×  10^{10}h^{-1}M_{\odot}$ and $10h^{-1}$kpc at  low redshift --
for the high redshift snapshots ($z>2$) a  value of $30h^{-1}$kpc is used. The
numerical  parameters were  set  to meet  the  requirements after  the
completion  of  many  tests  that  studied  the  convergence  for  the
correlation  function and  circular  velocities for  haloes and  their
subhaloes  \citep{klypin13}.  The  choice  of  parameters enables  the
simulation to  resolve well  the internal  structure of  all collapsed
systems  identified  with a  parallel  version  of the  bound  density
maximum (BDM) algorithm \citep{klypin97,riebe13},  thus, making it possible to
connect them with BOSS-like galaxies \citep{prada14,rodriguez15}.  From the cosmic
shear  point  of  view  this  allows us  to  measure  the  shear-shear
correlation function down  to small scales and  construct cosmic shear
power  spectrum that  is  very  accurate up  to  $l=10^4$  as will  be
shown. This  maximum limit, in  the angular  mode $l$, is  much larger
than  the value  expected to  be reachable  by the  future wide  field
surveys.

\begin{figure}
\includegraphics[width=\hsize]{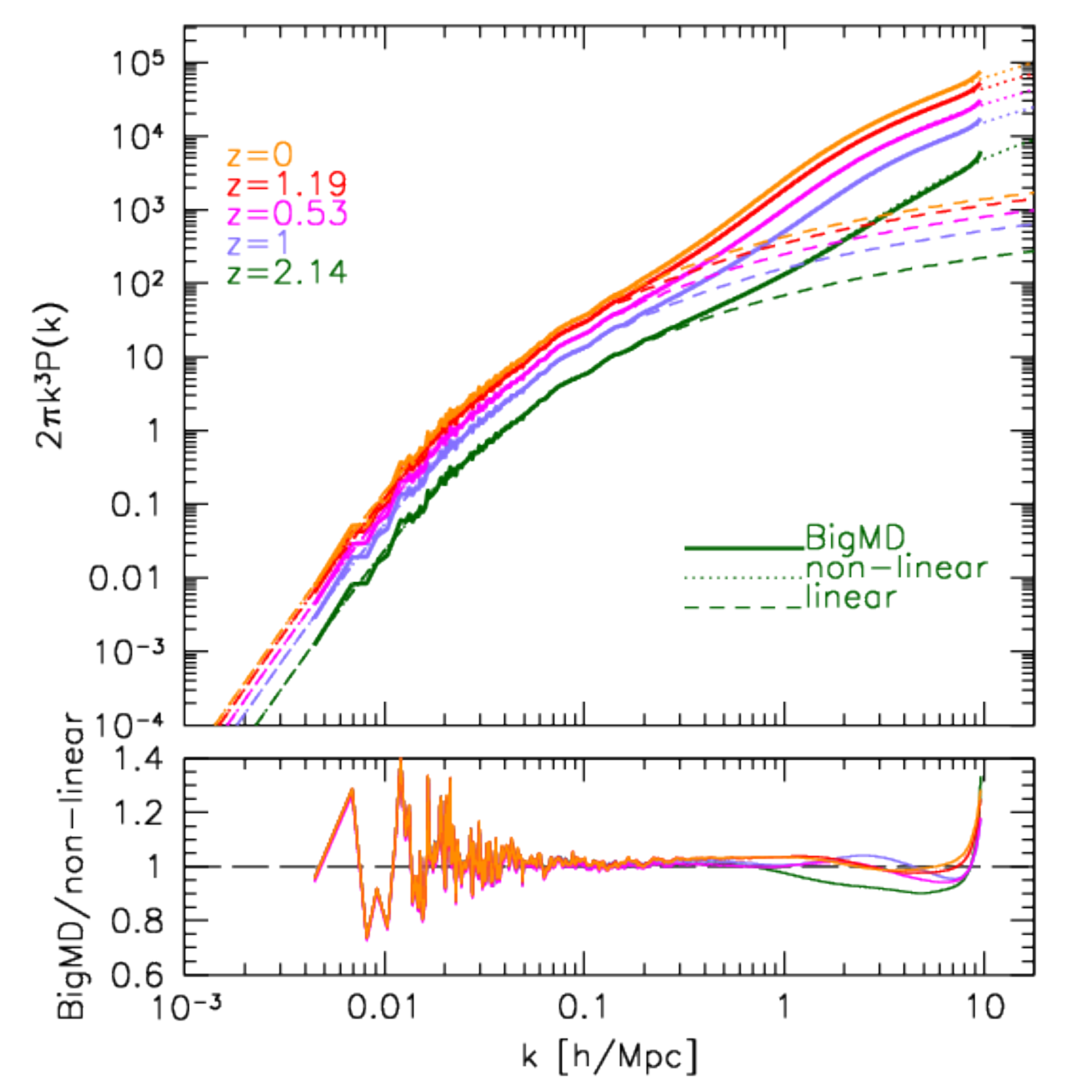}
\caption{Matter power spectra measured  at five different redshifts in
  the  simulation (solid  line).  The  dashed curves  show the  linear
  prediction  at  the  corresponding   redshifts  computed  using  the
  \textsc{CAMB}  code.    The  dotted   curves  show   the  non-linear
  predictions inputting the corresponding linear power spectrum in the
  extended     version      of     Halofit      \citep{smith03}     by
  \citet{takahashi12}. The  bottom panel  shows the ratio  between the
  non-linear power  spectrum measured  in the  simulation and  the one
  predicted by the analytical recipe.\label{figpk}}
\end{figure}

In Fig.~\ref{figpk} we show the matter power spectra at five different
redshifts  computed from  the  particle distribution  taken from  some
simulation snapshots (solid curves) as  indicated in the figure label.
Given  the large  amount  of  particles and  the  box-size, the  power
spectra  have been  computed on  a mesh  of $7680^3$  up to  a Nyquist
frequency  of $k=9.65$  $h/$Mpc  which correspond  to  650 kpc$/h$  in
comoving  scale. In  the same  figure linear  (dotted) and  non-linear
(dashed)  predictions  obtained  from \textsc{CAMB}  \citep{camb}  are
displayed.  For the non-liner matter  power spectrum model we adopt an
extended   version  of   the   Halofit   Model  \citep{smith03}   from
\citet{takahashi12} build on the  corresponding linear power spectrum.
The figure shows the very good agreement between the non-linear recipe
for the power spectrum and the  one measured in the simulation with an
uncertainty  well below  two  percents  for modes  between  0.1 and  1
$h$/Mpc. For  larger modes, below the  Nyquist value of the  mesh, the
deviations  from snapshots  with redshifts  $z>0$ reach  at most  four
percent with larger deviations at higher redshift.

\section{Methodology}
\label{secmethod}

In  this section  we describe  how the  lensing light-cones  have been
constructed   from   the   N-body  simulation   and   the   ray-tracing
procedure. The method used in this paper is similar to those presented
in \citet{vale03}  and \cite{hilbert09}.  The particles  are projected
onto  different lens  planes that  are distributed  along the  line of
sight    and     filling    the    light-cone    up     to    redshift
$z=2.3$. \cite{petkova14}  have studied  the impact  of the  number of
lens planes on lensing quantities such as the deflection angle and the
convergence. They  found that  the error in  the convergence,  even in
cases of strong  lensing, is less than 5\% if  planes are separated by
about 300 Mpc/$h$. In our case, we choose a distance between each lens
plane of  161 Mpc/$h$ which is  $N=24$ lens planes out  to 3.9 Gpc/$h$
comoving. In accordance  with this number, we select  24 snapshots out
of the 80  available in the simulation, for which  the redshift is the
closest to the estimated redshifts of the planes.

\begin{figure}
\includegraphics[width=\hsize]{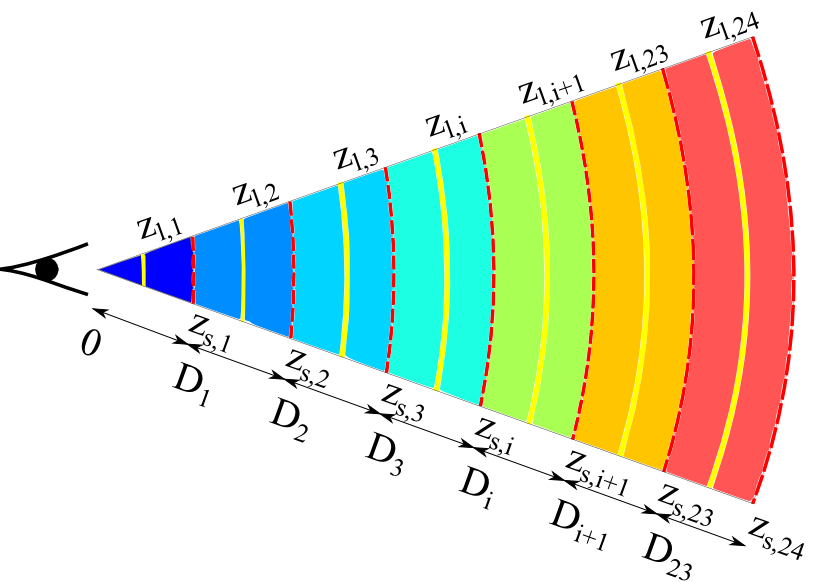}
\caption{Schematic representation of  the light-cone construction from
  the simulation.  The slices in colour  show the portion of the matter
  extracted from each snapshots with comoving distance between $D_{i}$
  and $D_{i+1}$, within the aperture of the field of view. Given $D_i$
  and $D_{i+1}$  we define  the lens redshift  at their  half distance
  $z_{l,i}=z[(D_i+D_{i+1})/2]$     and     the     source     redshift
  $z_{s,i}=z[D_{i+1}]$. \label{schemat}}
\end{figure}

\begin{figure*}
\includegraphics[width=\hsize]{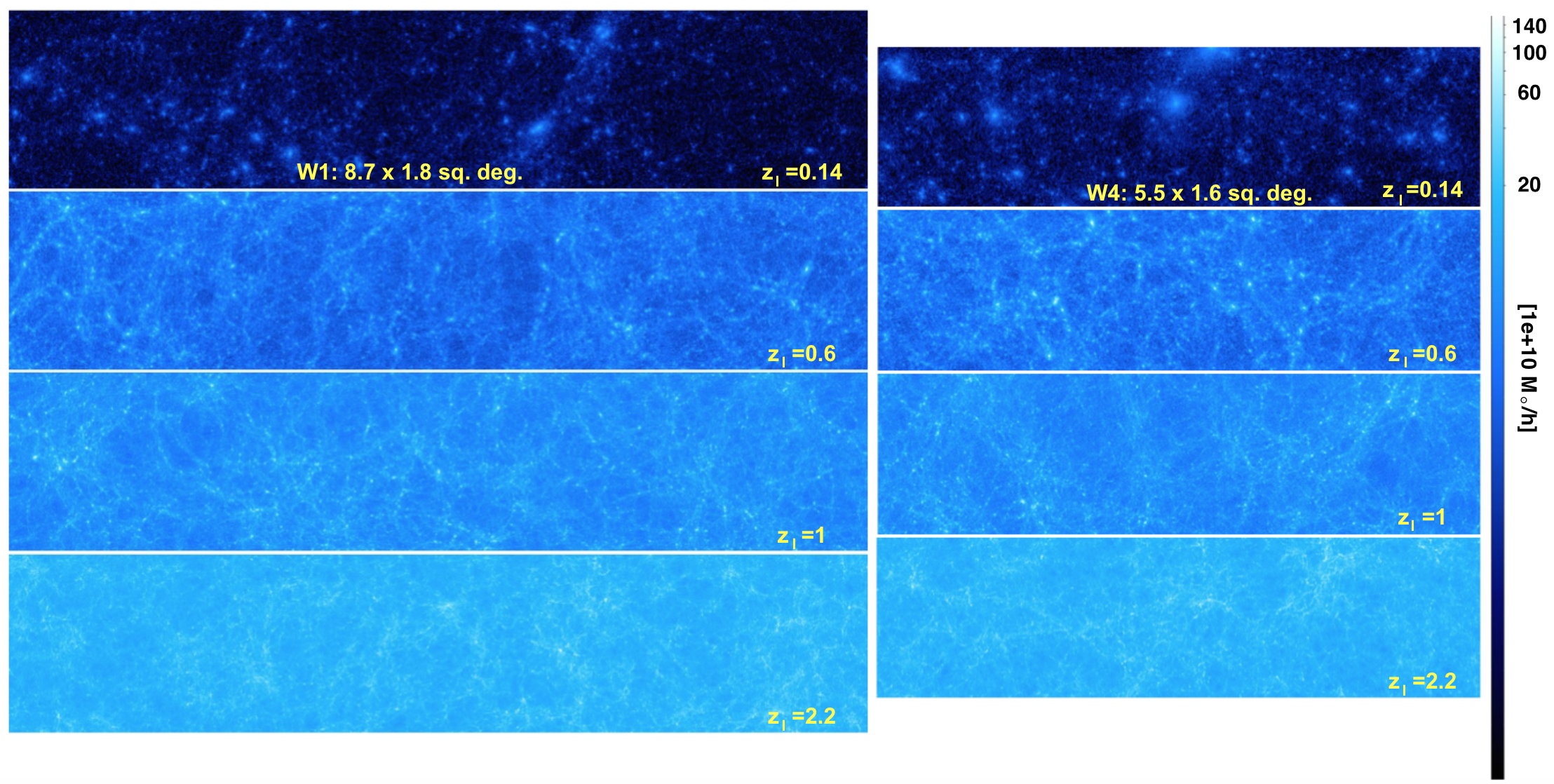}  
\caption{The two dimensional surface matter density distribution in four lens planes 
as extracted from the simulation: $z_l=$ 0.14, 0.6, 1 and
  2.2, from top to bottom respectively. While the left panels have the geometry
  of the W1 field, the right ones of W4; 
  with resolution $5220\times 1080$ and $3300\times 960$ pixels, respectively. 
  We remind the reader that the effective sizes of the two fields in the figure may 
  not be properly in scale with each other.  
  \label{w1and4lenses}}
\end{figure*}

\subsection{Light-cone geometry} 
\label{sectioncone}

For  the analyses  conducted  in  this work,  we  build two  different
light-cones geometries -- many other will be available in our database
-- one  covering a  region of  $8.7\times 1.8$  deg$^2$ and  another a
region of  $5.5\times 1.6$ deg$^2$,  that mimic  the W1 and  W4 fields
observed   by  the   VIPERS   extragalactic  survey   \citep{guzzo14},
respectively.   In  order  to   maximise  the  number  of  independent
light-cone realisations,  we remap  the 2.5$^3$  (Gpc/$h$)$^3$ cubical
volume           of           the           simulation           using
\textsc{BoxRemap}\footnote{\url{http://mwhite.berkeley.edu/BoxRemap}}
\citep{carlson10}.    \textsc{BoxRemap}   takes   advantage   of   the
periodicity of  the simulated box  to break  it into cells,  which are
then translated  by integer  offsets to  form cuboids.   The remapping
procedure keeps local structures intact.   We adjust the parameters in
order to  maximise the number of  light-cones that can be  embedded in
the  comoving  simulation  volume.   In this  adjustment,  we  avoided
configurations in which the original cube is remapped into a very long
and narrow  cuboid, where we found  that particles can be  lost or are
not  well  remapped.  For  the  W1  lightcones,  the original  box  is
remapped into a $3875\times 588\times 122$ Mpc/$h$ cuboid. For W4, the
shape is $3875\times 373\times 108$  Mpc/$h$.  From these cuboids, the
numbers of  independent realisations that  we were able to  create are
respectively 54 and 99 for the W1 and W4 fields.  Their properties are
summarised in Table~\ref{tab:fields}.

\begin{table}
\caption{Summary of simulated light-cones}
\begin{tabular}{ccc}
\hline
Field name & Size [deg$^2$] & \# Realisations  \\
\hline
W1 & $8.7×  1.8$ & 54\\
W4 & $5.5× 1.6$ & 99 \\
\hline
\end{tabular}
\label{tab:fields}
\end{table}

In  Fig.~\ref{schemat}  we  show   a  schematic  representation  of  a
light-cone construction. For each snapshot $i$, we remap its particles
to form  the cuboid shape given  above.  For each light-cone  and each
snapshot, we recenter  the particle positions according  to its centre
in the  cuboid and  convert its particle  positions from  Cartesian to
spherical coordinates. Particles with radial comoving distance between
$D_{i}$ and $D_{i+1}$ and inside the light-cone aperture are projected
onto   a  2D   angular  mesh   using  the   'cloud-in-cell'  technique
\cite{hockney88}. This procedure is repeated  for the M light-cones in
the  cuboid, thus  producing  M independent  lens  planes at  redshift
$z_{l,i}=z[(D_i+D_{i+1})/2]$ with $i \in [1,N]$.

In Fig.~\ref{w1and4lenses}  we show  the projected  mass per  pixel in
unit  of  $10^{10}M_{\odot}/h$,  distributed on  four  different  lens
planes  as indicated  in the  labels of  the figure.   Right and  left
panels refer to W1 and W4 geometry resolved with $5220\times 1080$ and
$3300\times 960$ pixels, respectively. The  effective sizes of the two
fields in the figure may not be properly in scale with each other.  No
truncation in the large scale structures is observed.

\subsection{Ray-Tracing through multiple planes}

\begin{figure}
\includegraphics[width=\hsize]{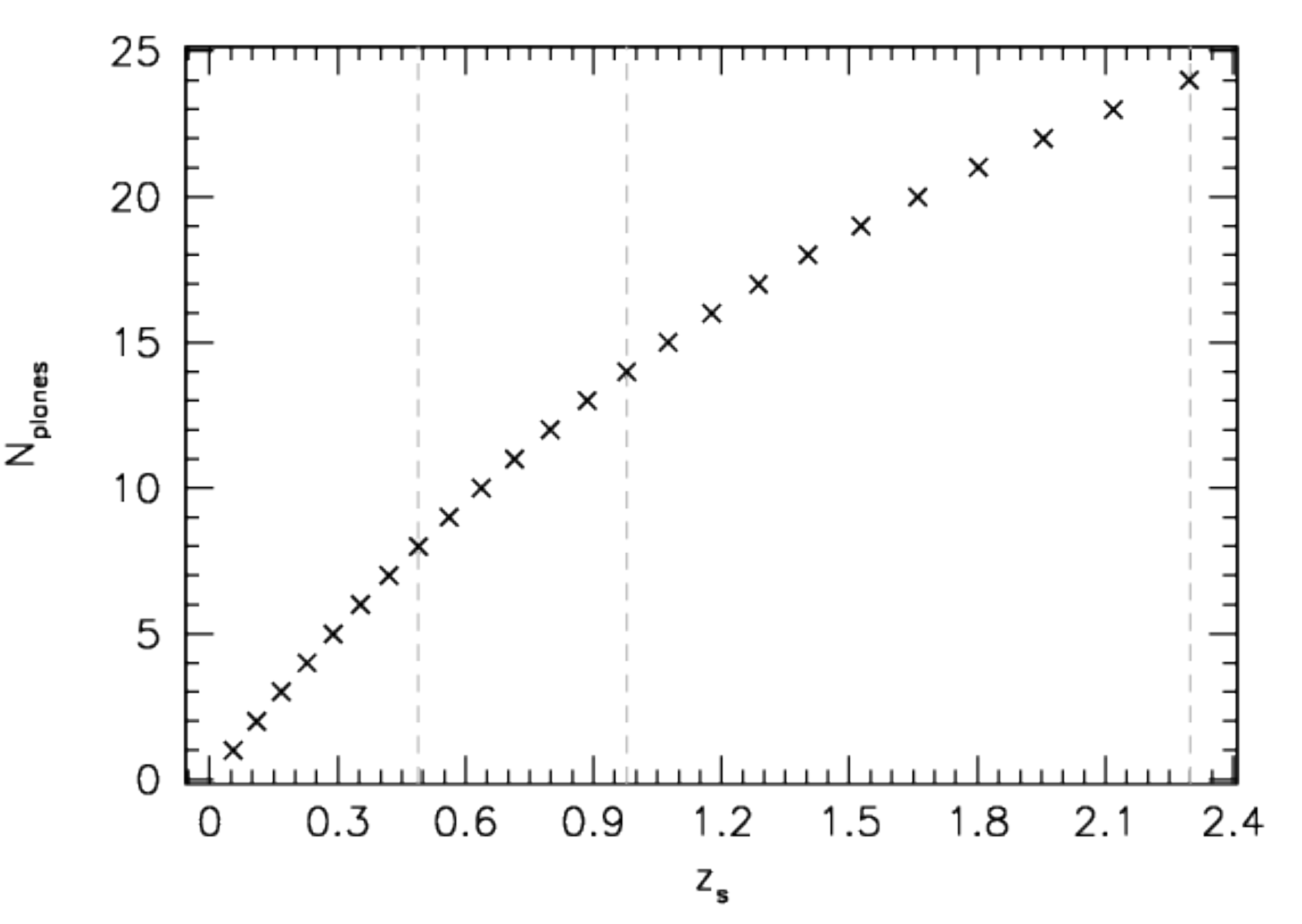}
\caption{Number  of  lens  planes  through which  the  ray-tracing  is
  performed as a function of  the source redshift. The vertical dashed
  lines  show as  a reference  $z_s=0.5$, 1  and 2.3  that we  will be
  considered as  reference in presenting some  lensing measurements in
  the following sections. \label{figplanes}}
\end{figure}

We  run  our  ray-tracing  code  \textsc{GLAMER}  on  each  of  the  M
realisations  of N  lens planes.  The ray-tracing  code is  run for  N
source redshifts in  the light-cone, i.e.  those  corresponding to the
boundaries  between 2  consecutive  snapshots $z_{si}  = D_{i+1}$.  In
Fig.~\ref{figplanes} we  show the  number of  planes through  which we
perform multi-plane ray-tracing  as a function of  the source redshift.
The vertical dashed  lines show $z_s=0.5$, 1 and 2.3,  that we will be
considered as reference in presenting some lensing measurements in the
following sections.   The particle  distributions from  the simulation
snapshots are  projected onto  different matter density  planes; these
are given  as input to \textsc{GLAMER}  \citep{metcalf14,petkova14} to
trace the light-rays  from different source redshifts  to the observer
-- although technically  the rays are shot the other  way around to be
sure that all rays intersect the the observer.

\begin{figure*}
\hspace{0cm}\includegraphics[width=\hsize]{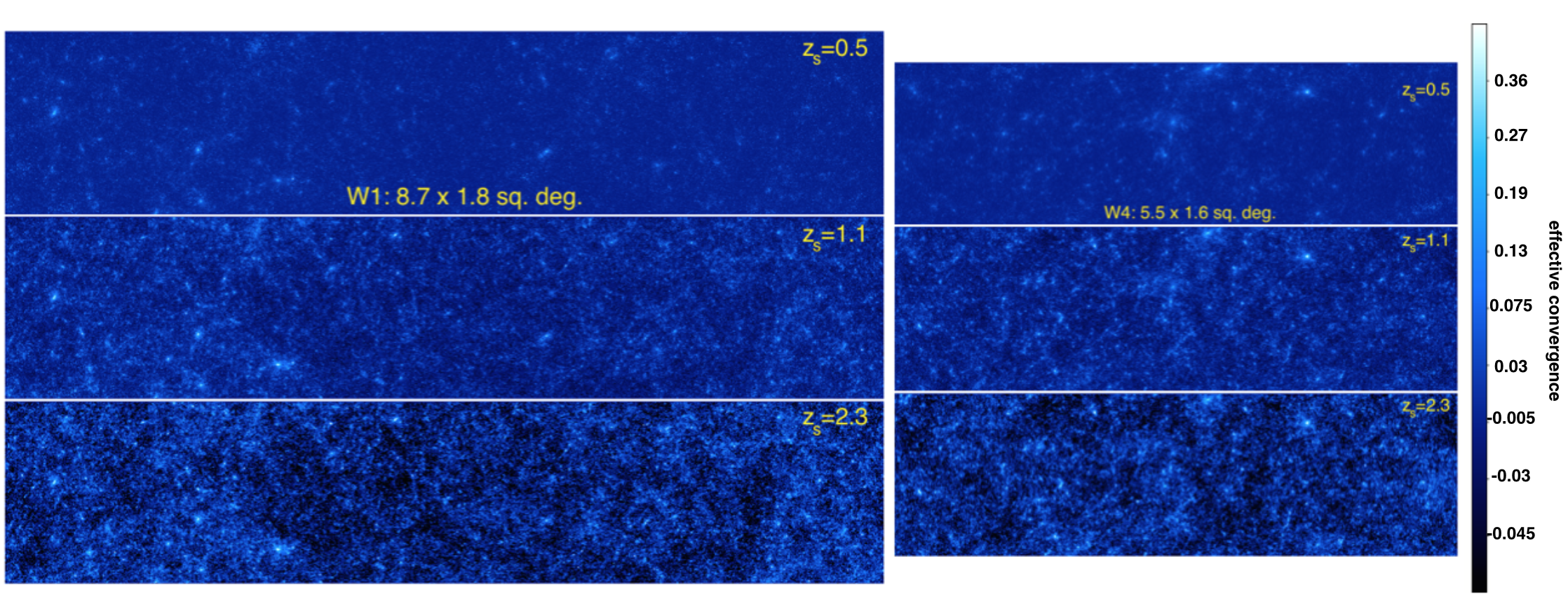}
\caption{Effective  convergence  maps of  one  realisation  of the  W1
  (left)  and W4  (right)  light-cones for  sources  located at  three
  different fixed redshifts. From top  to bottom we show the effective
  convergence  constructed from  our ray-tracing  pipeline considering
  all the  mass in the  light-cone up  to $z_s=0.5$, $1.1$  and $2.3$.
  The convergence  maps have the  same resolution of the  initial mass
  density  maps namely  corresponding to  an angular  resolution of  6
  arcsec per pixel.\label{figkappaw1andw4}}
\end{figure*}

A few definitions are required.  If the angular position on the sky is
${\pmb \theta}$ and  the position on the source plane  expressed as an
angle (the  unlensed position)  is ${\pmb  \beta}$, then  a distortion
matrix ${\bf A}$ can be defined as

\begin{align}
{\bf A} \equiv \frac{\partial {\pmb \beta}}{\partial {\pmb \theta} } = 
\left( 
\begin{array}{cc}
1-\kappa-\gamma_1 & \gamma_2 - \varsigma \\
\gamma_2 + \varsigma & 1-\kappa + \gamma_1
\end{array}
\right)\ .
\end{align}
The traditional decomposition of this  matrix is shown, where $\kappa$
is called  the convergence and  ${\pmb \gamma}$ represents  the shear.
The component $\varsigma$  is very small torsion which is  related to the
rotation  of the  image.  The  torsion is  of order  $\sim| \gamma|^2$
\citep{petkova14}  and  for  weak  lensing very  small,  but  will  be
retained here for completeness.

When there is a single lens plane, the convergence can be expressed as
a dimensionless surface density,
\begin{equation}\label{eq:magnification_matrix}
\kappa({\pmb \theta}) = \frac{\Sigma({\pmb \theta})}{\Sigma_{\rm
    crit}}\ ,
\end{equation}
where 
\begin{equation}
\Sigma_{\rm crit} \equiv \dfrac{c^2}{4 \pi G} \dfrac{D_l}{D_s D_{ls}}
\end{equation}
is called  the critical  density, $c$  is the speed  of light,  $G$ is
Newton's  constant  and  $D_l$  $D_s$ and  $D_{ls}$  are  the  angular
diameter   distances   between  observer-lens,   observer-source   and
source-lens,  respectively.  In  general, with  multiple lens  planes,
this is not the case however.

The deflection caused by a lens  plane, ${\pmb \alpha}$, is related to
the  surface density  on  the plane,  $\Sigma({\bf  x})$, through  the
differential equations
\begin{align}\label{eq:single_plane_kappa}
\nabla^2 \phi({\bf x}) = \frac{4 \pi G}{c^2} \Sigma({\bf x})   ~~~,~~~
  {\pmb \alpha}({\bf x}) = \nabla  \phi({\bf x}) .
\end{align}
where the  derivatives are with  respect to  the position on  the lens
plane.  These equations are solved  on each source plane by performing
a Discrete Fourier Transform (DFT)  on the density map, multiplying by
the appropriate factors and then transforming back to get a deflection
map with  the same resolution as  the density map.  With  the same DFT
method the  shear caused by  each plane is  simultaneously calculated.
In order to  reduce boundary effects, due  to non-periodic conditions,
before  going  in  the  Fourier  space the  maps  are  enclosed  in  a
zero-paddling region.  Because the  rays are propagated between planes
assuming a uniformly distribution of  matter, the matter that has been
projected onto  the planes over-counts  the mass in the  universe.  To
correct  for  this the  ensemble  average  density  on each  plane  is
subtracted.  This  way each plane has  zero total mass on  average and
the average redshift-distance relation is preserved.

After the deflection  and shear maps on each plane  are calculated the
light-rays are traced  from the observers through the  lens planes out
to the  desired source redshift.   The shear and convergence  are also
propagated     through      the     planes     as      detailed     in
\cite{petkova14}\footnote{It was found the the exact method derived in
  \cite{petkova14}  was  prone to  numerical  errors  so it  has  been
  somewhat   modified  in   the   current  version   of  the   code.}.
\textsc{GLAMER} preforms a complete ray-tracing calculation that takes
into account non-linear  coupling terms between the planes  as well as
correlations between  the deflection and  the shear.  No  weak lensing
assumption is made at this stage.  The rays are shot in a grid pattern
with the same resolution as the mass maps.

In the  Figures \ref{figkappaw1andw4} we display  the convergence maps
for one  realisation of the W1  (left) and W4 (right)  fields at three
different source redshifts, increasing from top to bottom as indicated
in the caption. We emphasise  that in doing the multi-plane ray-tracing
we  have followed  the ray  bundles through  $N=24$ lens  plane up  to
redshift $2.3$, $15$ up to redshift $1$ and $8$ up to $z=0.5$.

\subsection{CFHTLS source redshift distribution}
\begin{figure}
\includegraphics[width=\hsize]{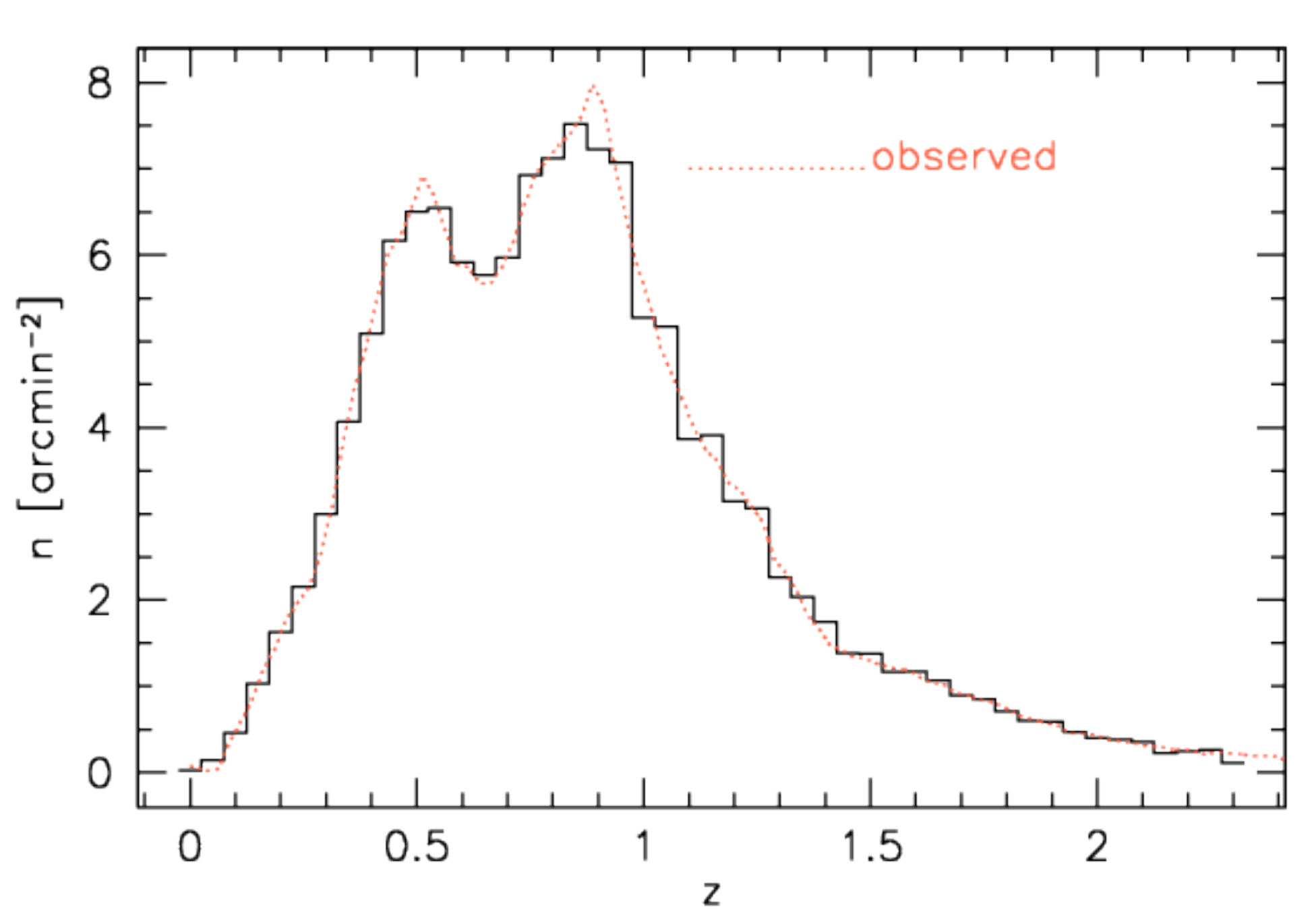}
\caption{Source  redshift distribution  adopted to  compute the  shear
  catalogue from our light-cone realisations. The dashed curve represents
  the CFHTLS source redshift  distribution, while the solid histogram
  refers the one obtained  from one realisation of the W1 field, randomly distributing
  the sources in the field-of-view.\label{fignz}}
\end{figure}

Another important ingredient that needs  to be taken into account when
producing  lensing simulations  is  the redshift  distribution of  the
sources.  Here  we consider the CFHTLens  source redshift distribution
\citep{hildebrandt12}  as computed  within the  W1 and  W4 fields.  In
Fig.~\ref{fignz}  we   show  with   the  dashed  curve   the  redshift
distribution of  the sources observed  in the two fields  by CHFTLens,
for comparison the solid histogram displays the distribution extracted
from one realisation  of the W1 field. Since the  impact of the source
galaxy clustering on lensing statistics  is beyond the purpose of this
first work,  the location of the  sources within the field  of view is
performed  randomly.  The  tool for  extracting the  lensing catalogue
with a desired source redshift distribution and a generated light-cone
is also available  in our database.  Upon request, users  can also ask
for  an  effective  convergence  map according  to  a  desired  source
redshift distribution table.

Once the redshift and the source positions within the field of view are known, 
we compute the corresponding lensing properties -- convergence and shear -- linearly 
interpolating the quantities in the field and between different planes. In this way
we extract self-consistent catalogs for the desired field of view according 
to a specific source redshift distribution\footnote{The catalogs are publicly available 
at \url{https://bolognalensfactory.wordpress.com/home-2/multdarklens}}. 
For each  field of view, we randomly draw eight catalogues, hence in total 
432 and 891 lensing catalogues for W1 and W4, respectively. 

In  our  online achieve  we  provide  a tool  that  allows  a user  to
calculate shear  and convergence  maps and  catalogs with  any desired
source redshift distribution.

\section{Cosmic-Shear and Lensing Signals} 
\label{sectionstat}

The convergence power spectrum, to first order, can be expressed as an
integral of  the 3D  matter power spectrum  computed from  the present
time up to the source  redshift \citep{bartelmann01}.  In this section
we review the calculation of  the cosmic shear power spectrum adopting
the Born approximation (light rays travel along unperturbed paths) and
the weak lensing proximation (ignore all terms higher than first order
in $\kappa$ and  $\gamma$).  Since our ray-tracing  simulations do not
make these approximations we can check their validity by comparing the
two methods of calculating the shear power spectrum.

Following an  unperturbed light-ray through inhomogeneous  universe it
is possible  to calculate  the first  order convergence,  $\kappa$, in
terms        of         the        matter         over        density,
$\delta =  (\rho -  \overline{\rho})/\overline{\rho}$, that  it passes
through
\begin{equation}
\kappa(w_s,\pmb{\theta}) = \dfrac{3 H_0 \Omega_m}{2 c^2} \int_0^{w_s} \mathrm{d}w
\dfrac{f(w)f(w_s-w)}{f(w_s)} \dfrac{\delta(f(w)\pmb{\theta},w)}{a(w)}
\end{equation}
where $f(w)$  represents the radial  function:
\begin{equation}
 f(w) = \begin{cases} 
             K^{-1/2} \sin\left(K^{1/2}w\right) & \;\;\; (K>0)\\
              w & \;\;\; (K=0) \\
              (-K)^{-1/2} \sinh\left[ \left(-K \right)^{1/2} w\right] & \;\;\; (K<0) \\
            \end{cases}
\end{equation}
depending on whether the curvature of the universe K is positive, zero
or negative; and $a=1/(1+z)$ the  scale factor.  The convergence power
spectrum for sources at a fixed redshift is:
\begin{equation}
 \langle \hat{\kappa}(\pmb{l}) \hat{\kappa}*(\pmb{l}') \rangle = \left(2 \pi\right)^2 \delta_D(\pmb{l}-\pmb{l}') P_{\kappa}(l)
\end{equation}
with
\begin{equation}
P_{\kappa}(l) = \dfrac{9 H_0^4 \Omega_m^2}{4 c^4} \int_0^{w_s} 
\mathrm{d}w \dfrac{f^2(w_s-w)}{f^2(w_s) a^2(w)} P_{\delta} \left( \dfrac{l}{f(w)},w\right)\,.
\label{eqpowerkappa}
\end{equation}
\citep{Kais98,Kais92}   Considering  a   normalised  source   redshift
distribution  $p(z_s)\mathrm{d}z_s=g(w_s)\mathrm{d}w_s$  we can  write
down  the associated  convergence power  spectrum of  a population  of
galaxies as
\begin{equation}
P_{\kappa}(l) = \dfrac{9 H_0^4 \Omega_m^2}{4 c^4} \int_0^{w_s} 
\mathrm{d}w \dfrac{W^2(w)}{a^2(w)} P_{\delta} \left( \dfrac{l}{f(w)},w\right)\,,
\end{equation}
where $W(w)$ represents the weighted source redshift distribution:
\begin{equation}
W(w) = \int_w^{\infty} \mathrm{d} w' g(w') \dfrac{f(w'-w)}{f(w')}\,.
\end{equation}
In the same way we can obtain an estimate of the effective convergence
map   given  a   source   redshift  distribution   by  weighting   the
contributions of the different constructed planes.

In real space, the direct measurement of weak lensing is the two point
shear correlation functions  $\xi_+$ and $\xi_-$ that  can be obtained
from galaxy  ellipticity measurements $\epsilon_t$  and $\epsilon_{×}$
(aligned  and at  45 degrees  to  the line  connecting to  galaxies)--
tangential  and cross  component,  respectively --  by averaging  over
galaxy pairs with angular distance $|\pmb{\theta}_i - \pmb{\theta}_j|$
in a bin $\theta$:
\begin{equation}
  \xi_{±}(\theta) = \dfrac{\sum_{ij} \omega_i \omega_j \left[ \epsilon_t(\pmb{\theta}_i)
      \epsilon_t(\pmb{\theta}_j) ± \epsilon_{×}(\pmb{\theta}_i) \epsilon_{×}(\pmb{\theta}_j) \right]}{\sum_{ij}\omega_i \omega_j}\,
\end{equation}
\citep{schneider02} where $\omega$ represents the weight obtained from
galaxy  shape measurement  pipeline. The  two point  shear correlation
functions can  be related to  the cosmic  shear power spectrum  by the
following relation:
\begin{equation}
\xi_{+/-} = \dfrac{1}{2 \pi} \int_0^{\infty} \mathrm{d} l ~l
P_{\kappa}(l) J_{0/4}(l\theta)\,,
\end{equation}
where $J_0$  and $J_4$ are  the Bessel functions  and we have  set the
B-mode power spectrum equal to zero in agreement with the weak lensing
prediction.

It is  also common to  measure the  variance of the  convergence field
filtered with window function or aperture
\begin{eqnarray}
M\left(\theta\right) = \int_0^\theta d^2\theta' ~W(|\theta'|) \kappa(\pmb{\theta}') 
\end{eqnarray}
with the normalisation
\begin{eqnarray}
\int_0^\theta d\theta~ W(\theta) = 0.
\end{eqnarray}
The convergence $\kappa$  is not directly measured in  a shear survey,
but \cite{1996MNRAS.283..837S}  showed that  $M\left(\theta\right)$ is
equivalent to
\begin{eqnarray}
M\left(\theta\right) = \int_0^\theta d^2\theta' ~Q(|\theta'|) \gamma_t(\pmb{\theta}') 
\end{eqnarray}
with
\begin{eqnarray}
Q\left(\theta\right) = \frac{2}{\theta^2} \int_0^\theta d\theta' ~ \theta' W(\theta')  - W(\theta)
\end{eqnarray}
which can be  measured from the tangential  ellipticities of galaxies.
We   will    investigate   a   few   choices    for   $W(\theta)$   in
section~\ref{sectionstat}.

\begin{figure*}
\includegraphics[width=\hsize]{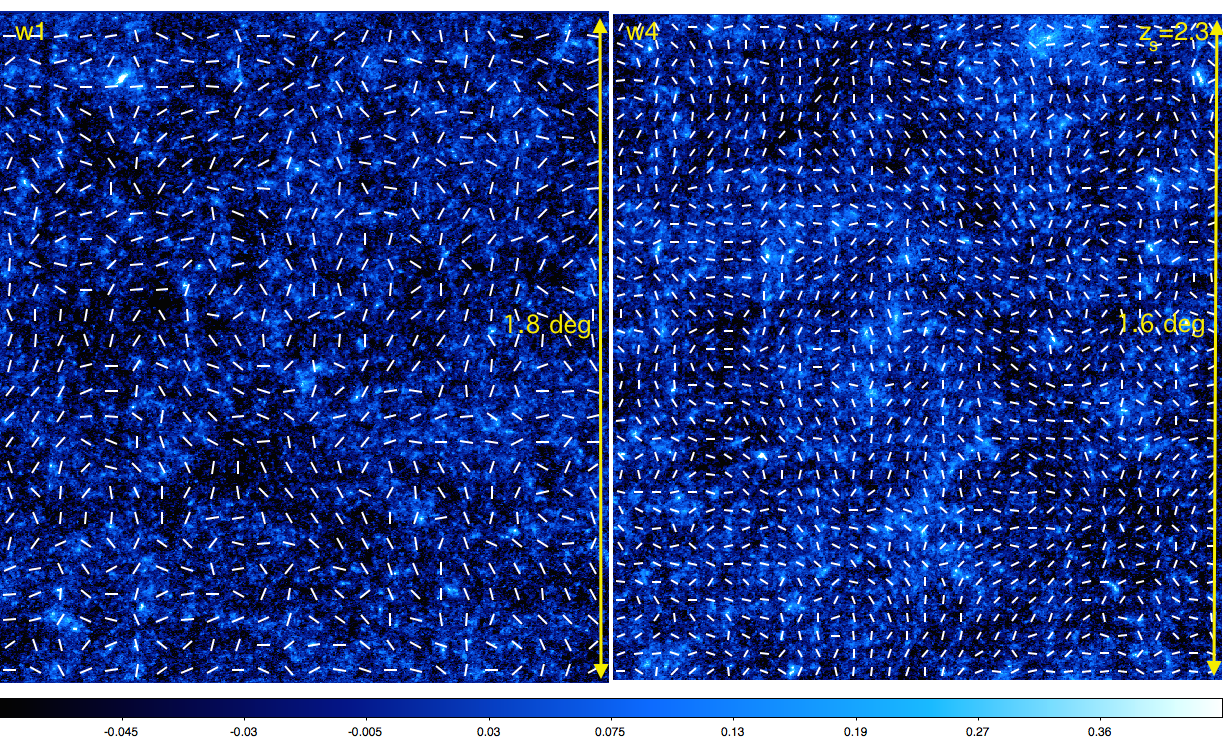}
\caption{Effective convergence maps for  sources at redshift $z_s=2.3$
  for a square region the W1 (left) and the W4 (right) fields. In this
  case we show sticks -- that in each panel have the same angular size
  --   representing  the   direction   of   the  corresponding   shear
  field.\label{figkstick}}
\end{figure*}

In Fig.~\ref{figkstick} we show the  effective convergence maps of two
cropped regions in the  W1 and W4 fields of view with  1.8 deg and 1.6
deg on a  side, respectively, pixels have a resolution  of $6$ arcsec.
The convergence maps have been  constructed considering all the matter
density distribution in  the light-cone up to  redshift $z_s=2.3$.  In
each  squared  panel  the  sticks  represent  the  directions  of  the
corresponding  shear   field,  which  are   consistently  tangentially
directed around large density peaks.

\begin{figure}
\includegraphics[width=\hsize]{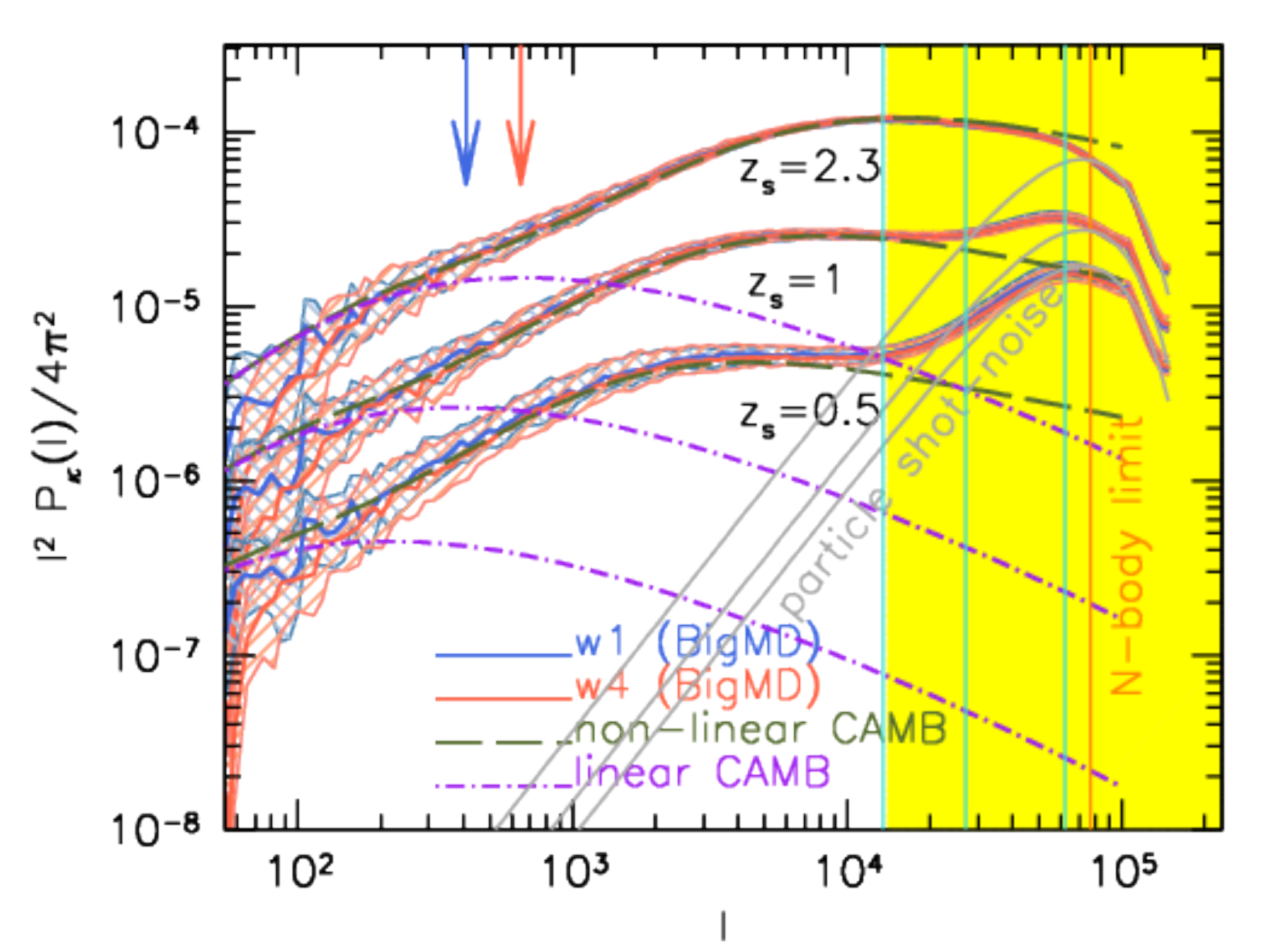}
\caption{Cosmic-shear power spectrum from  the simulation (thick solid
  curve)  considering  sources  located  at  different  redshifts,  as
  indicated in the label. The  cosmic shear power spectrum predictions
  from the  linear and the  non-linear Halo-Fit matter  power spectrum
  are  also shown  in the  figure  with dashed  and dot-dashed  curve,
  respectively.  The  two coloured  vertical arrows indicate,  for the
  corresponding  colour type,  the  multipole scale  which sample  the
  largest coordinate within the field of view. The three cyan vertical
  lines (from  left to right)  indicate the angular scales  mode above
  which  the power  spectrum is  dominated by  particle shot-noise  at
  source redshift $z_s=0.5$, $1$  and $2.3$, respectively.  The yellow
  shaded   area  indicates   the  region,   starting  from   the  mode
  corresponding  to $96$  arcsec, below  which any  of the  considered
  source redshift  maps are not  affected by particle  shot-noise. The
  grey curves show  the particle shot-noise contribution  at the three
  considered source redshifts.\label{figpk1}}
\end{figure}

\begin{figure}
\includegraphics[width=\hsize]{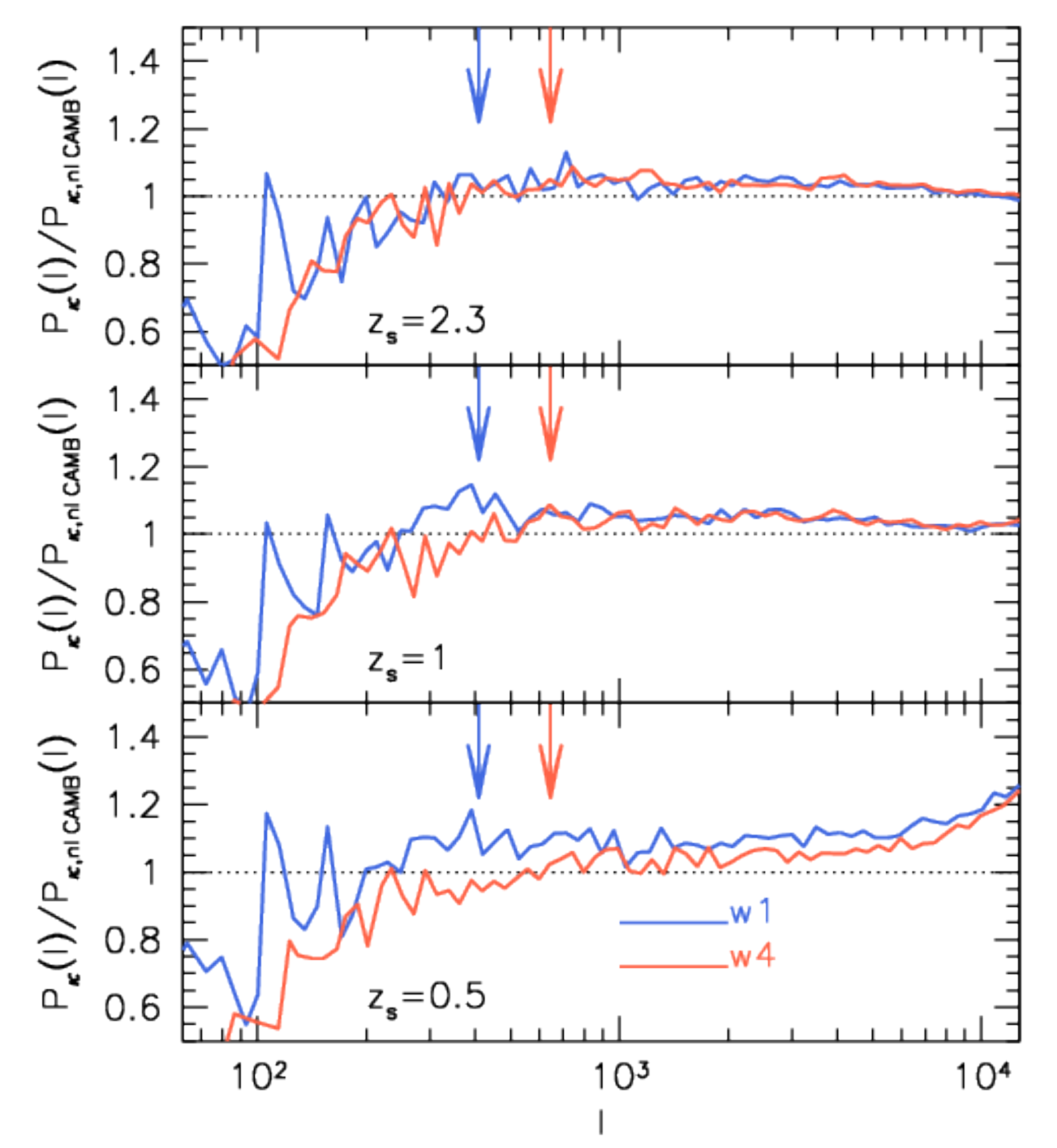}
\caption{Ratio between the measured cosmic  shear power spectra in the
  simulated fields  of view  and the  non-linear predictions by \citep{takahashi12} at three
  different redshifts. The two  coloured vertical  arrows indicate,  for the
  corresponding  color  type, the  multipole  scale  which sample  the
  largest coordinate within the field of view in the W1 and W4 fields.
  \label{figpk2}}
\end{figure}

In Fig.~\ref{figpk1}  we show the convergence  power spectrum computed
for  three  different  fixed  source  redshifts from  the  54  and  99
realisations of  the W1 and  W4 fields, respectively.  The  solid blue
and  red  curves represent  the  median  on the  different  light-cone
realisations, while the corresponding shaded region encloses the first
and  the third  quartile of  the distribution  at fixed  $l$. The  two
coloured vertical  arrows indicate, for the  corresponding color type,
the multipole scale  which sample the largest coordinate  of the field
of view.   The three  cyan vertical lines  indicate the  minimum scale
(from left  to right)  up to which  the power spectra  of the  maps at
redshift  $z_s=0.5$,  $1$  and  $2.3$ are  not  affected  by  particle
shot-noise.  The orange  vertical line -- labelled as  N-body limit --
corresponds to $l_N=1/\sigma_N$ with $\sigma_N = 0.05 N_{part}^{-1/3}$
and  $N_{part} =  3840^3$, as  discussed also  by \citet{vale03}.  The
yellow shaded area indicates the region below which any the considered
source  redshift  maps are  not  affected  by particle  shot-noise  --
starting from  the mode corresponding  to $96$ arcsec.  We  remind the
reader that  the minimum  scale where particle  noise start  to become
worthy of consideration  depends on the source redshift and  is of the
order of $24$ arcsec for $z_s=2.3$.  The grey curves show the particle
noise contribution  \citep{vale03} at  the three  corresponding source
redshifts that can be read as:
\begin{equation}
l ^2 P_{\kappa,\,\textrm{SN}}(l) \propto \dfrac{l^2}{N_{part}^3} \exp\left(-l^2/l^2_{SN}\right)\,
\end{equation}
with $l_{SN} = 2 \pi/(3 \theta_{pix})$ ($\theta_{pix}=6\,\textrm{arcsec}$).
 
The  dashed green and the dot-dashed magenta  curves show the
convergence  power  spectra computed  by  integrating  the linear  and
non-linear power  spectra from  \textsc{CAMB} up to  the corresponding
considered  source  redshift.  From  the  figure  we notice  that  the
agreement  between the  power  spectrum computed  from the  ray-traced
field  of view  and  from  non-linear CAMB  agree  quite  well with  a
difference of the order of only few percents.  This is more evident in
Fig.~\ref{figpk2} where  we present the ratio  between the convergence
power  spectrum measured  in the  two  fields W1  and W4  and the  one
computed using  equation (\ref{eqpowerkappa}) adopting  the non-linear
\textsc{CAMB}  matter  power  spectrum.   The figure  shows  that  the
agreements between  the theoretical modelling  of the cosmic  shear and
the shear measured from ray-tracing  simulations agree within $5\%$ up
a scale of  $l\approx 3000$ which represents the  largest multipole up
to  which future  extragalactic surveys  are expected  to measure  the
lensing  power spectrum.   Small differences  between simulations  and
non-linear  predictions  may be  due  to  non-linear lensing  effects,
cosmic variance,  or to the fact  that the non-linear modeling  of the
power spectrum by \citep{takahashi12} does not quite capture the small
scale clustering  of the matter  in the numerical simulation.   In the
figure it can  be seen that the  limited size of the field  of view is
responsible  for the  drop in  the  signal at  small multiples  (large
scales)  which gives  an idea  of the  accuracy with  which the  power
spectrum normalisation can be measured given  the size of the field of
view.

\begin{figure}
\includegraphics[width=\hsize]{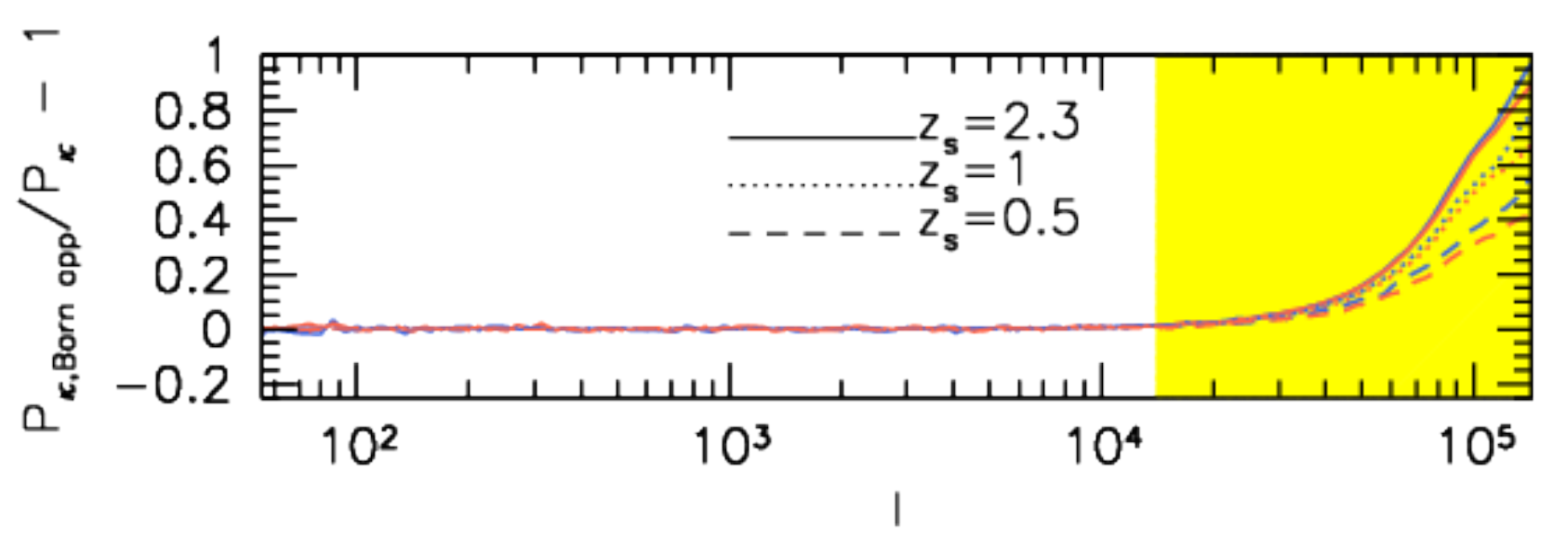}
\caption{Relative  residuals  between  the  median  convergence  power
  spectra computed from  the different realisations of  the two fields
  assuming  or  discarding  the  Born  approximation,  for  the  three
  considered source redshifts. As in the previous figures blue and red
  colors refer  to the W1 and  the W4 field, respectively.  The yellow
  shaded area is like in Figure \ref{figpk1}. \label{figpkbApp}}
\end{figure}

Differences between analytic models and simulations may be also due to
the fact that the analytic  calculations are made under the assumption
of the  Born approximation, where  the ray paths approximated  as what
they  would  be in  a  homogeneous  metric, and  discarding  lens-lens
coupling.  To quantify  this we re-run \textsc{GLAMER} again on  all the fields
and  realizations  turning  off  the Born  approximation.   In  Figure
\ref{figpkbApp}  we show  the  relative residuals  between the  median
convergence power spectra computed for the two fields with and without
the  Born approximation,  for  three different  source redshifts.   As
noticed  by  \citet{schaefer12}  performing an  analytic  perturbative
expansion in  the light  path the Born  approximation is  an excellent
approximation for weak cosmic lensing  but fails at small scales where
strong and quasi-strong lensing takes place. From the figure we notice
also  that the  relative difference  at  small scales  depends on  the
number of  lens planes  considered in  the ray-tracing  and so  on the
source redshift.

\subsubsection{Power Spectrum Correlation Matrix}

To accurately measure the  cosmological  parameters  from  the lensing
power spectra of the W1 and  W4 light-cones, the cross-correlation
between measurements of the power at different scales must be known.
This can be quantified by the power spectrum correlation matrix that 
is mainly influenced by the specific survey geometry and the non-Gaussian 
nature of the density distribution. 

From the  different light-cones realisations  we  build  up  the  covariance
matrix from the definition:
\begin{equation}
M(l,l') = \langle P_{\kappa}(l) - \bar{P}_{\kappa}(l)\rangle\langle P_{\kappa}(l') - \bar{P}_{\kappa}(l')\rangle
\end{equation}
where $\langle \bar{P}_{\kappa}(l)\rangle$ represents the best estimate of the
power  spectrum  at the  mode  $l$  obtained  from the median of  all  the
corresponding light-cone realisations and  $P_{\kappa}(l)$ represents the measurement of one realisation. 
The matrix is then normalised as follows: 
\begin{equation}
m(l,l') = \dfrac{M(l,l')}{\sqrt{M(l,l) M(l',l')}}\,.
\end{equation}
The covariance matrix constructed in this  way accounts both for a Gaussian
and non-Gaussian  contribution arising from  mode coupling due  to
non-linear clustering  and for the survey geometry
\citep{scoccimarro99,cooray01,harnois-deraps12,sato13}.  Off-diagonal
terms with value near unity indicate high correlation  while values approaching zero
indicate no correlation.

\begin{figure*}
\centering
\hspace{-0.5cm}
\includegraphics[width=0.32\hsize,angle=0]{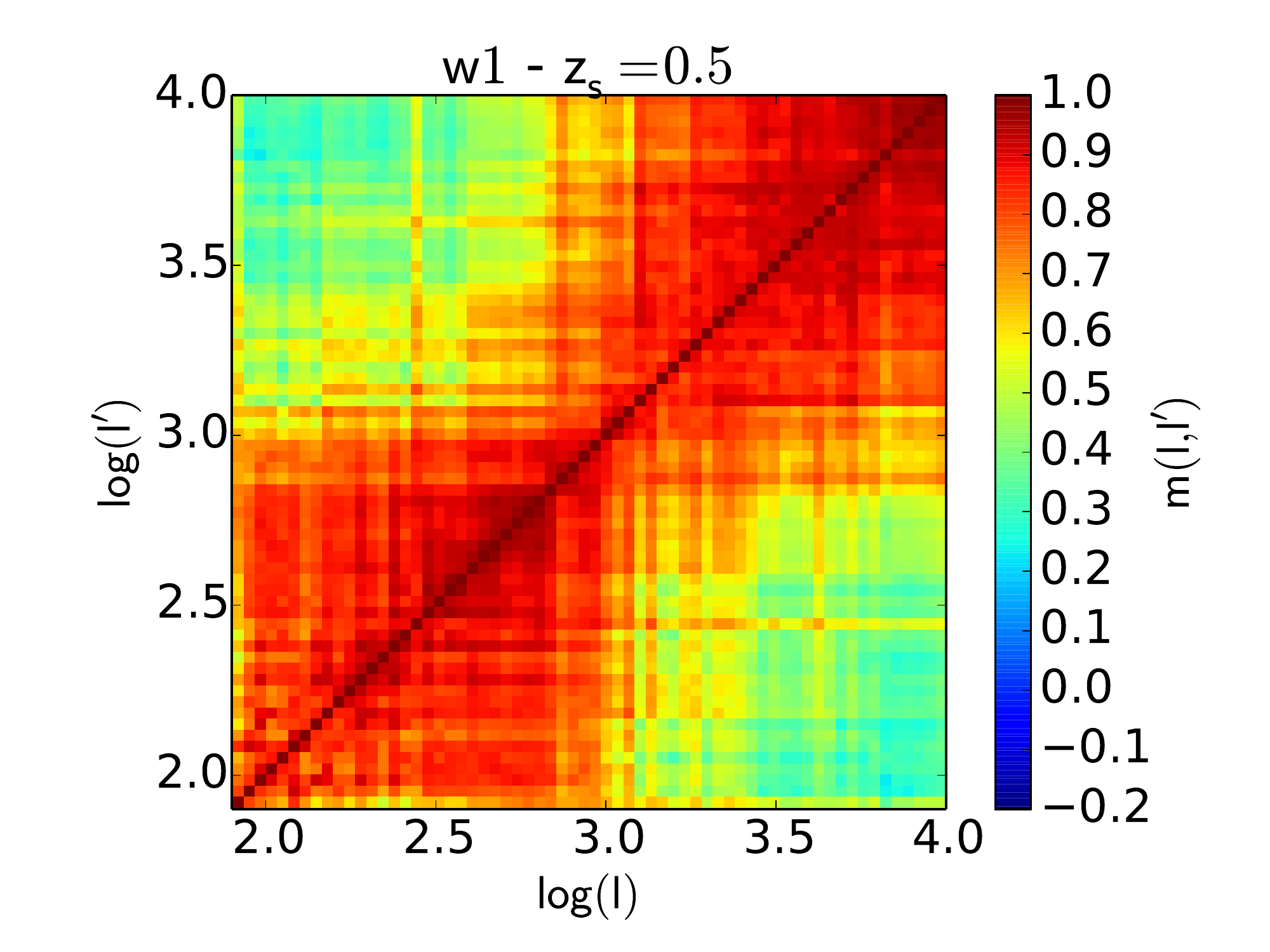}
 \hspace{-0.2cm}
\includegraphics[width=0.32\hsize,angle=0]{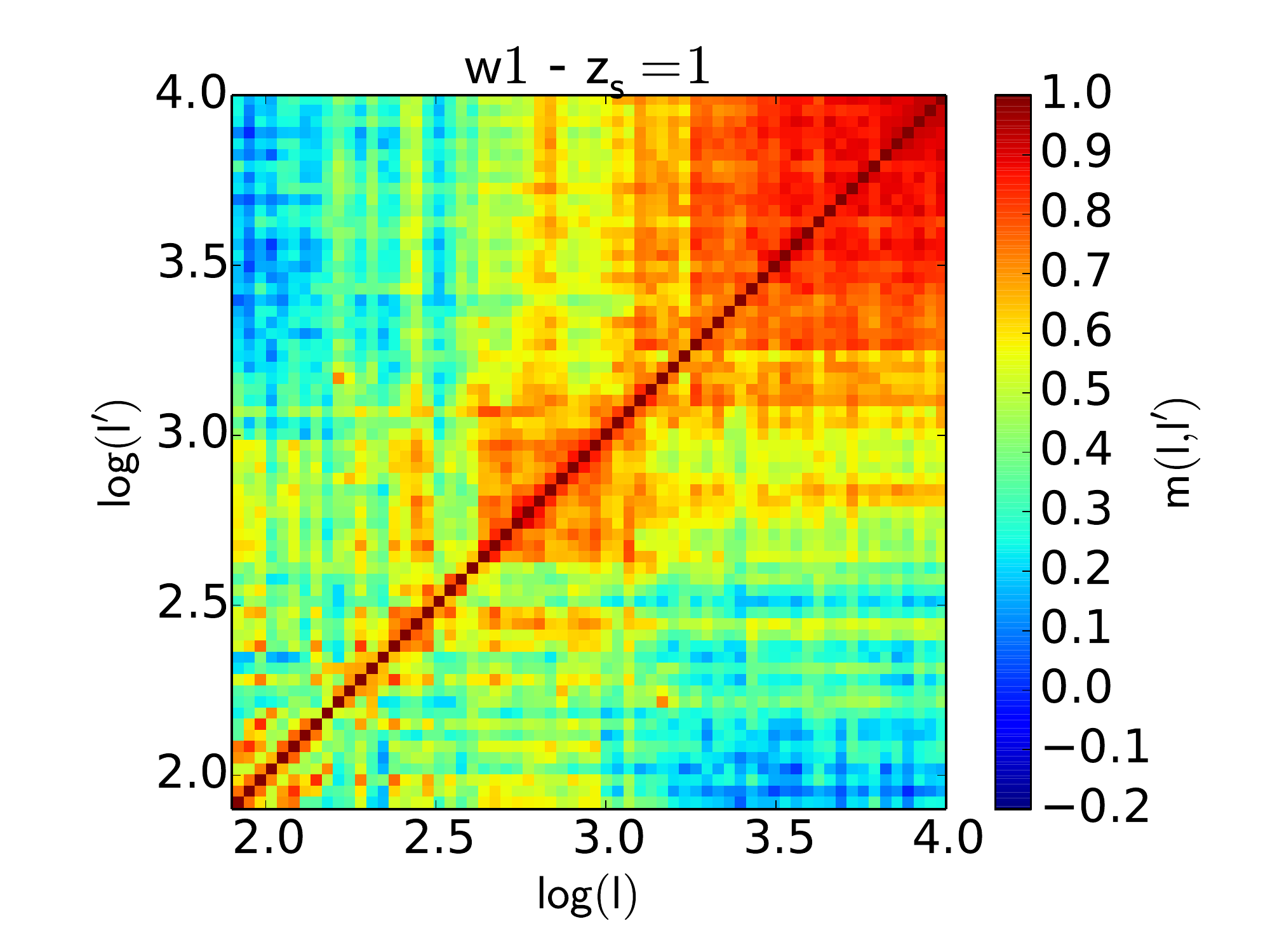} 
 \hspace{-0.2cm}
\includegraphics[width=0.32\hsize,angle=0]{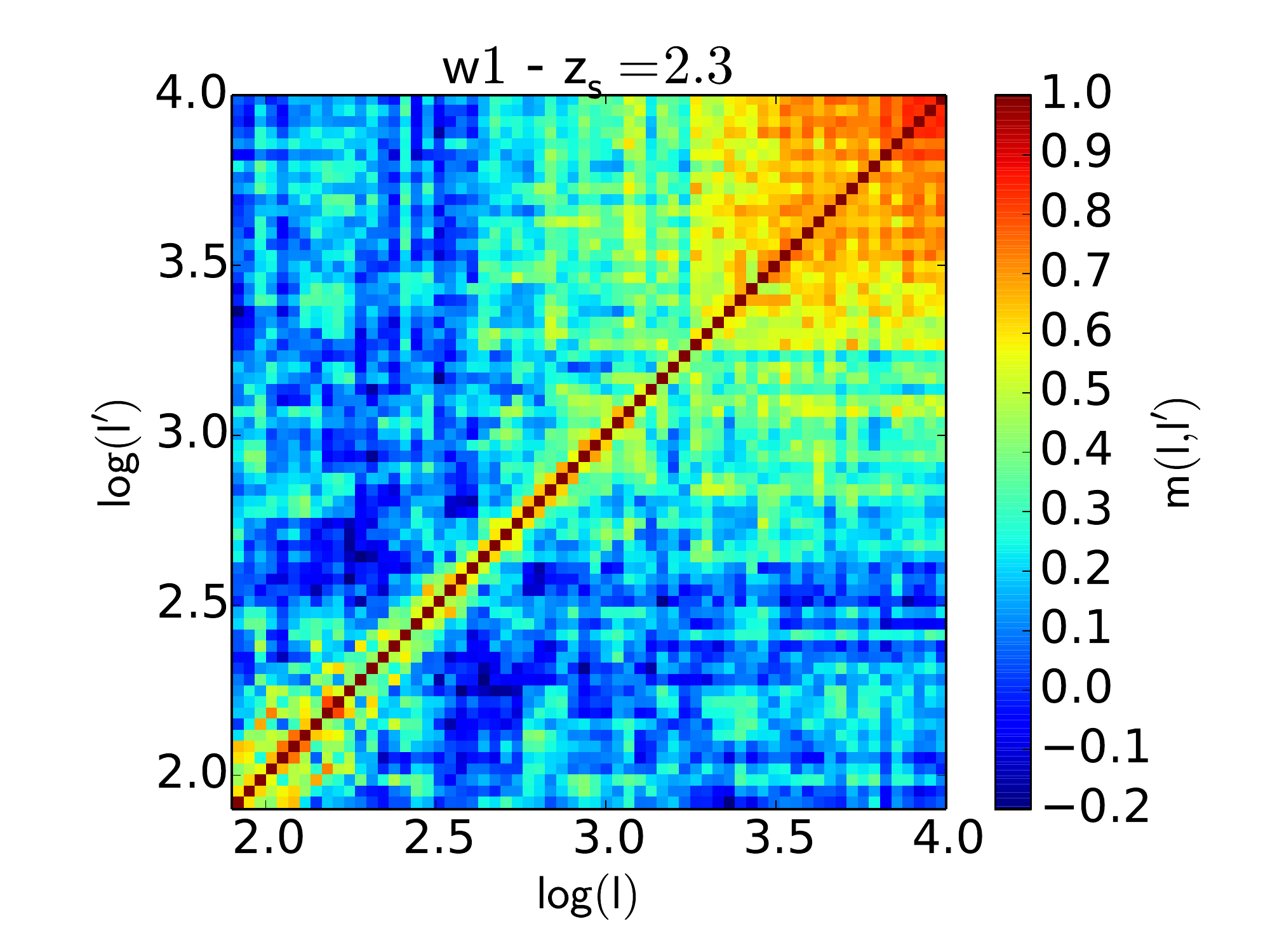} \\
\hspace{-0.5cm}
\includegraphics[width=0.32\hsize,angle=0]{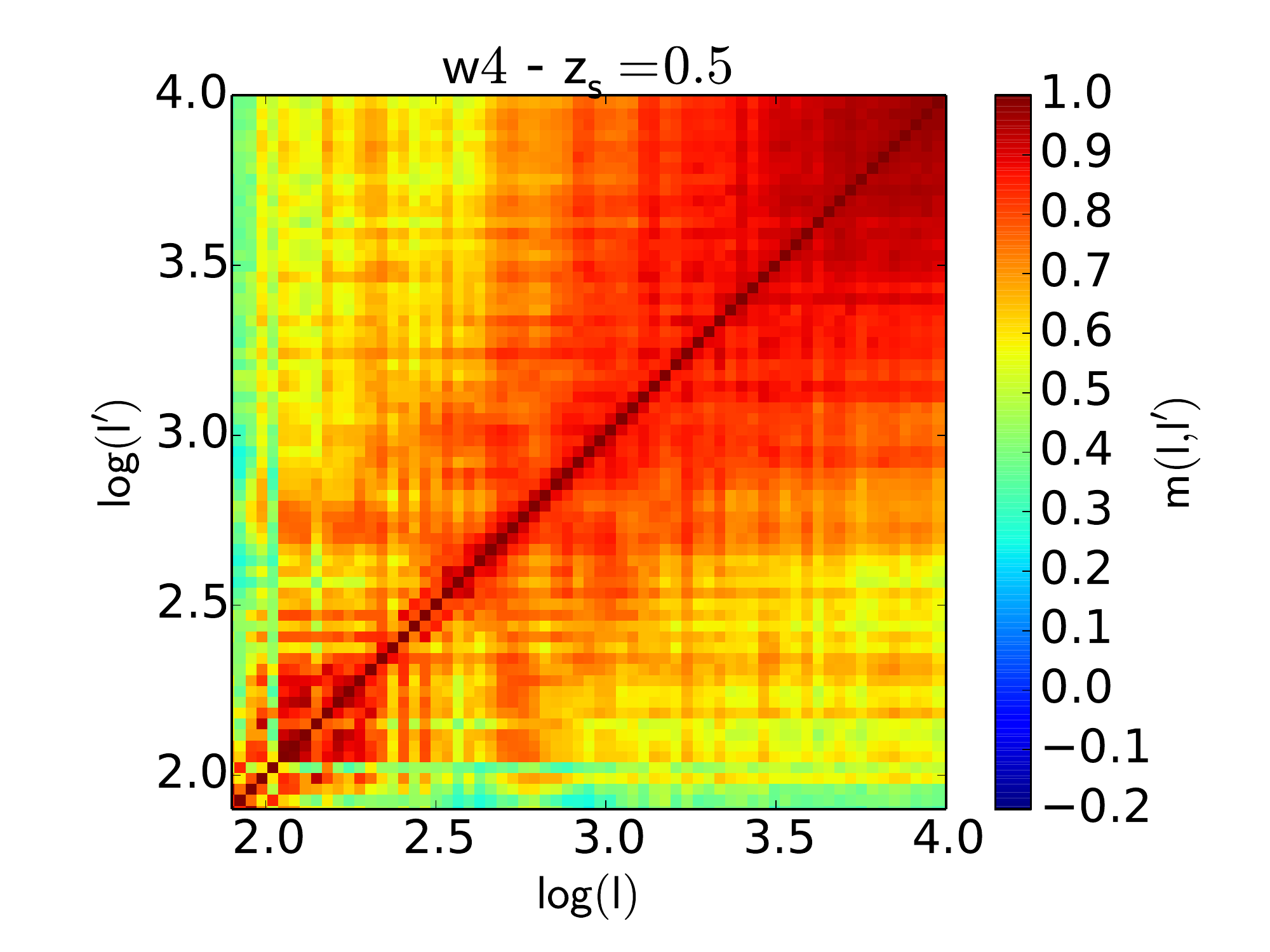}
 \hspace{-0.2cm}
\includegraphics[width=0.32\hsize,angle=0]{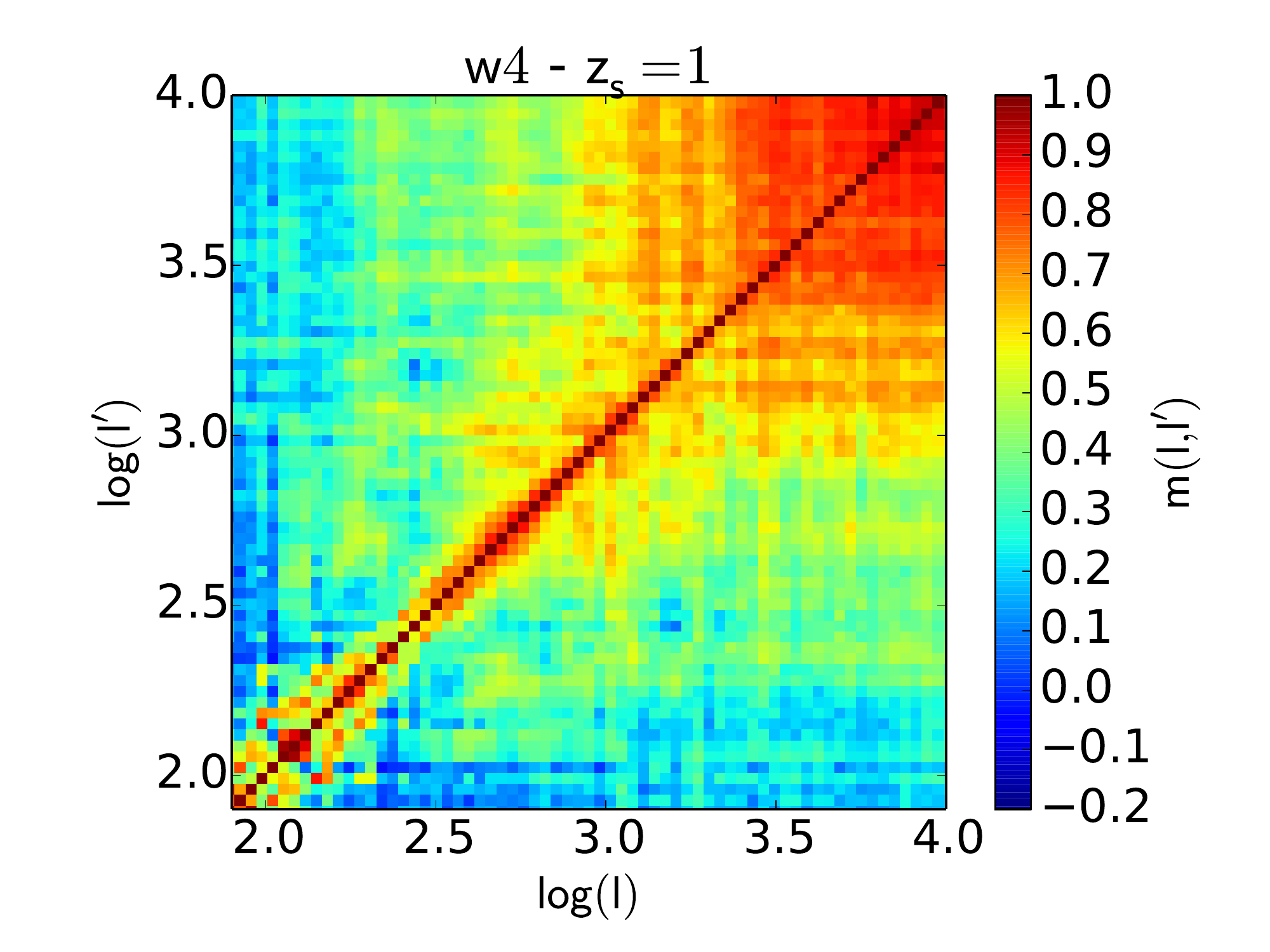} 
 \hspace{-0.2cm}
\includegraphics[width=0.32\hsize,angle=0]{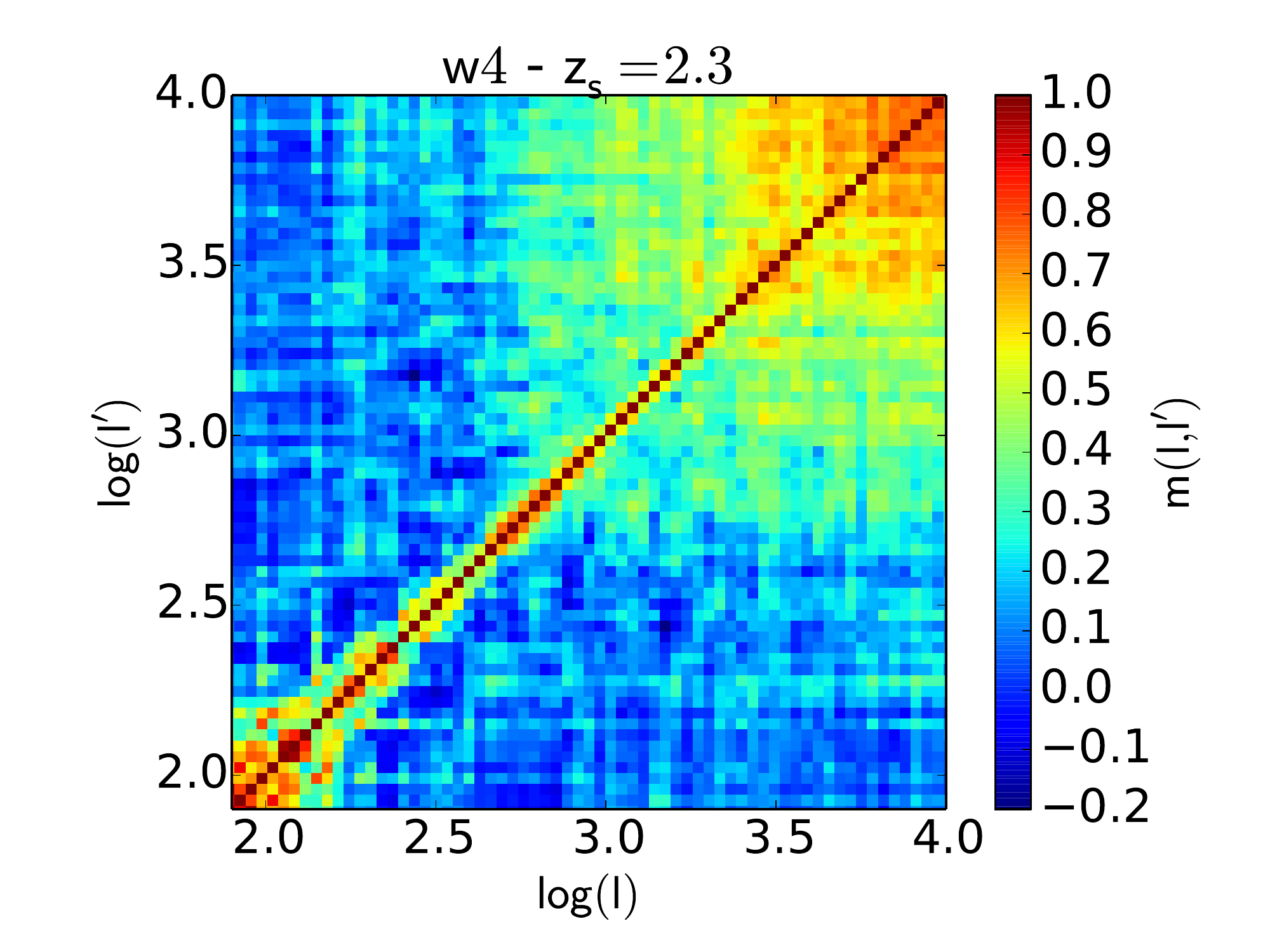}
\caption{Normalised  cosmic  shear  covariance matrices  for  the  two
  fields  of view  -- W1 in  the top  panels  and W4  in the  bottom,
  respectively. As indicated in the label the matrices are presented considering 
  sources at three different fixed source redshifts.\label{figmatrices}}
\end{figure*}

In Fig.~\ref{figmatrices}  we show the normalised  covariance matrices
for the W1 (top) and the W4 (bottom) fields of view assuming $\Delta l
\approx 0.075$.  We  again show the cases where sources  are fixed and
located at three  different redshifts, increasing from  left to right,
as indicated in the label.  It  is interesting that at small redshifts
the correlation of the power  spectrum between two different $l$-modes
is  stronger than  at higher  redshifts.  For  sources with  $z_s=2.3$
correlations are present  only at small scales, large  $l$ modes.  The
enhancement  of the  correlation at  large $l$  (small scales)  at low
redshift  is intrinsic  to  the non-Gaussian  statistics  of the  halo
clustering.   From the  figure we  also notice  that the  low redshift
covariance matrices depend  on the considered field  of view.  Because
W1 is a longer stripe the enhancement in the correlation occur already
at larger scales with respect to W4.

\subsubsection{one-point probability distributions}

\begin{figure*}
\includegraphics[width=0.325\hsize]{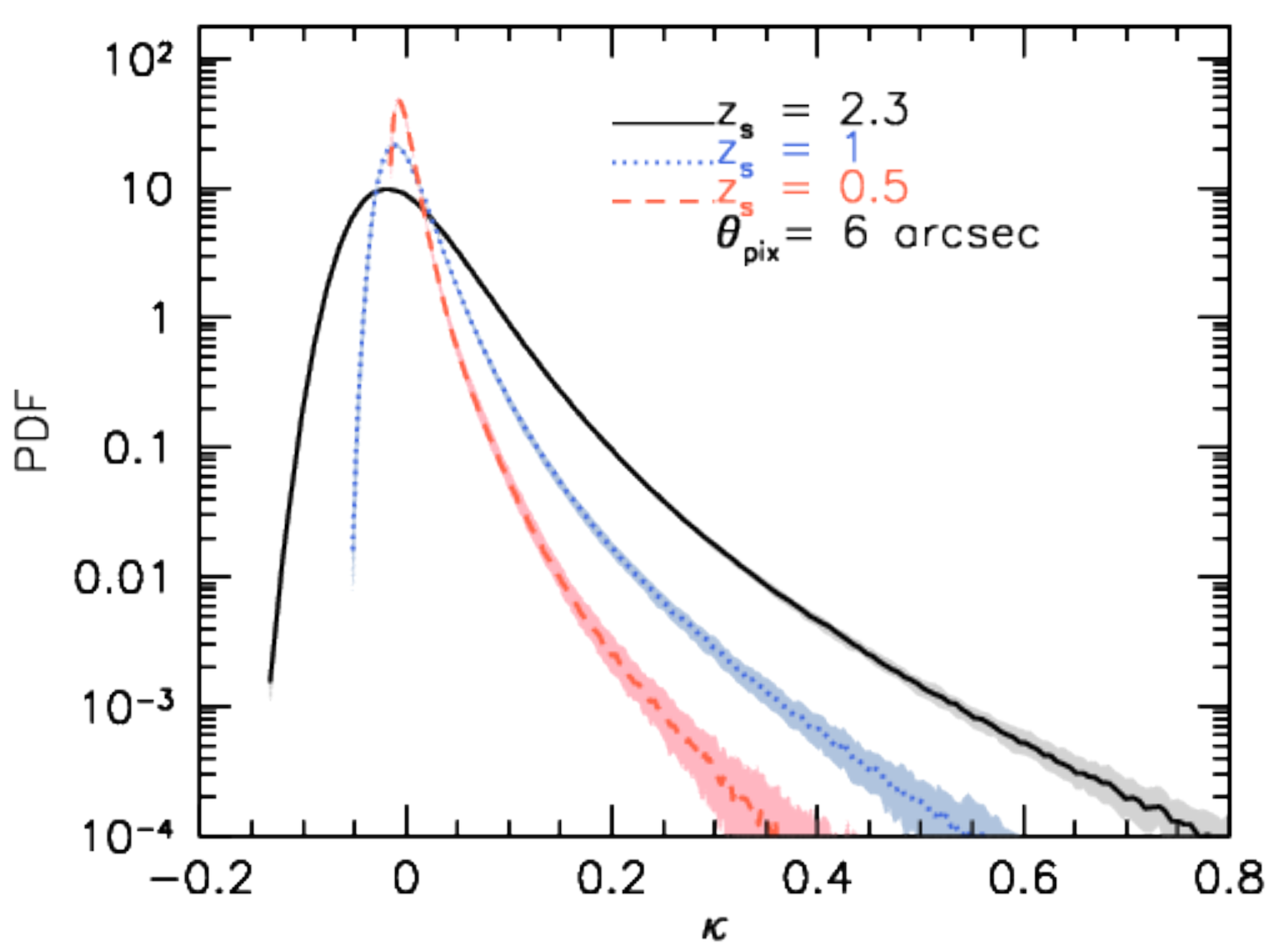}
\includegraphics[width=0.325\hsize]{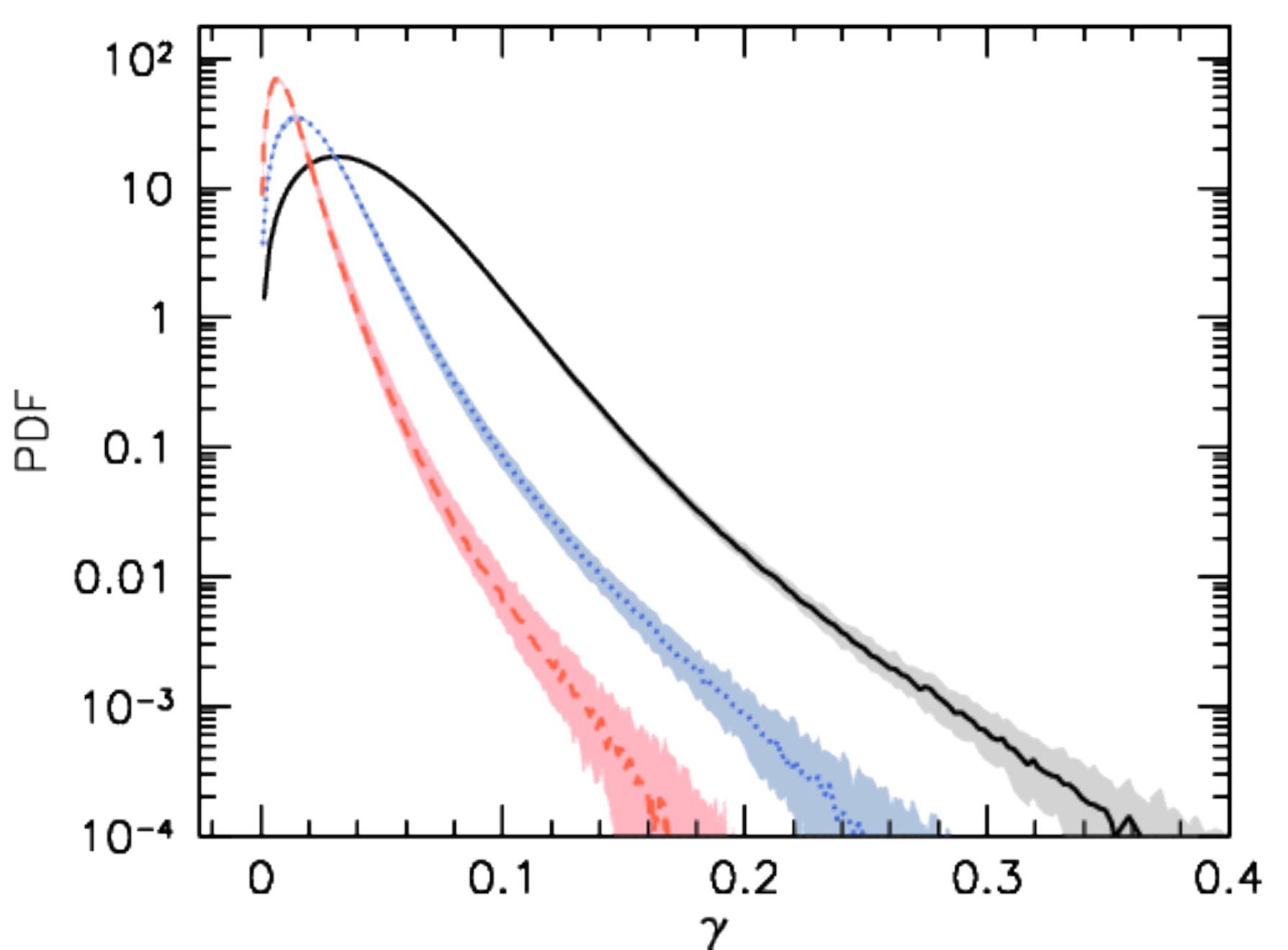}
\includegraphics[width=0.325\hsize]{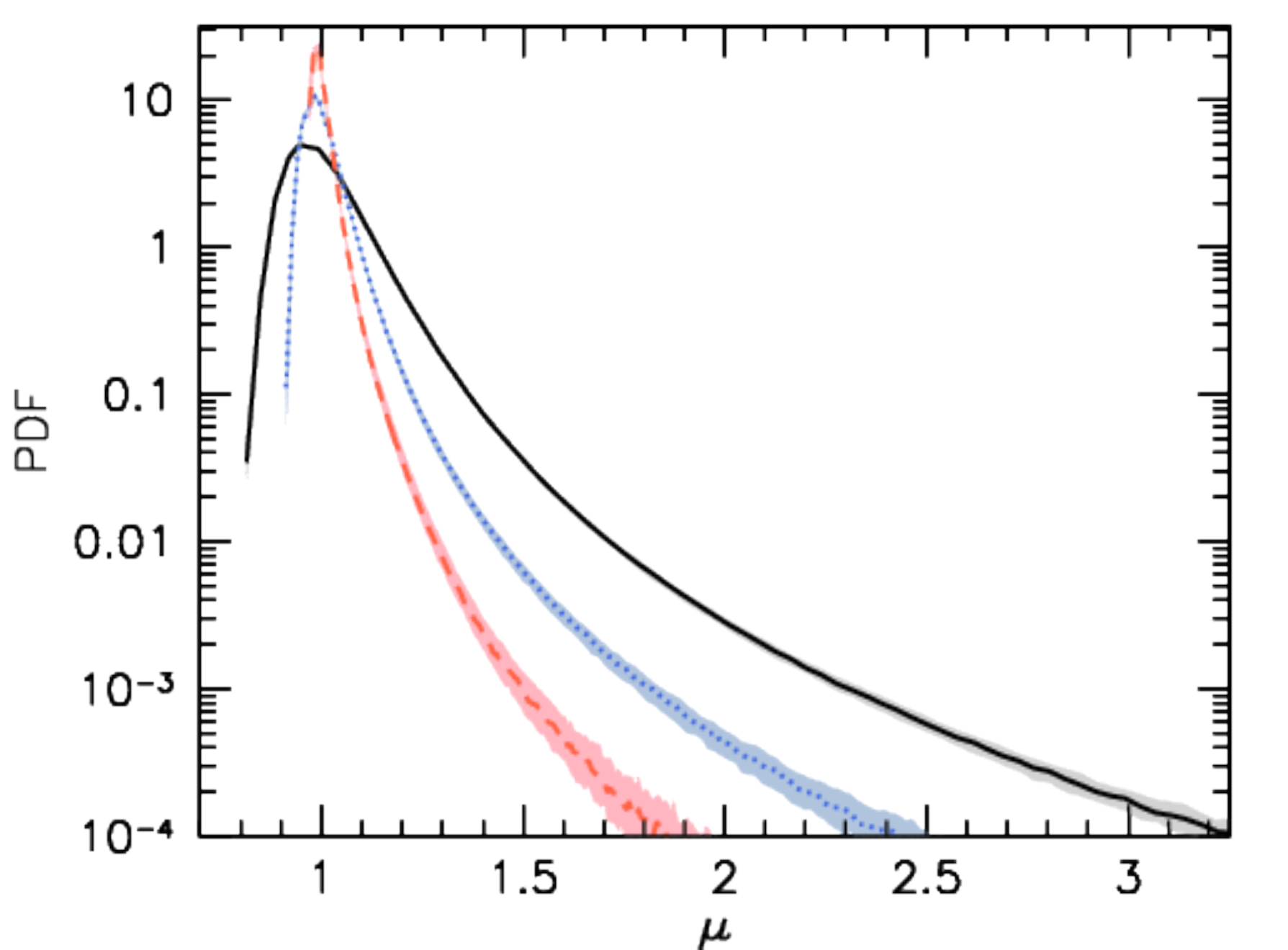}
\includegraphics[width=0.325\hsize]{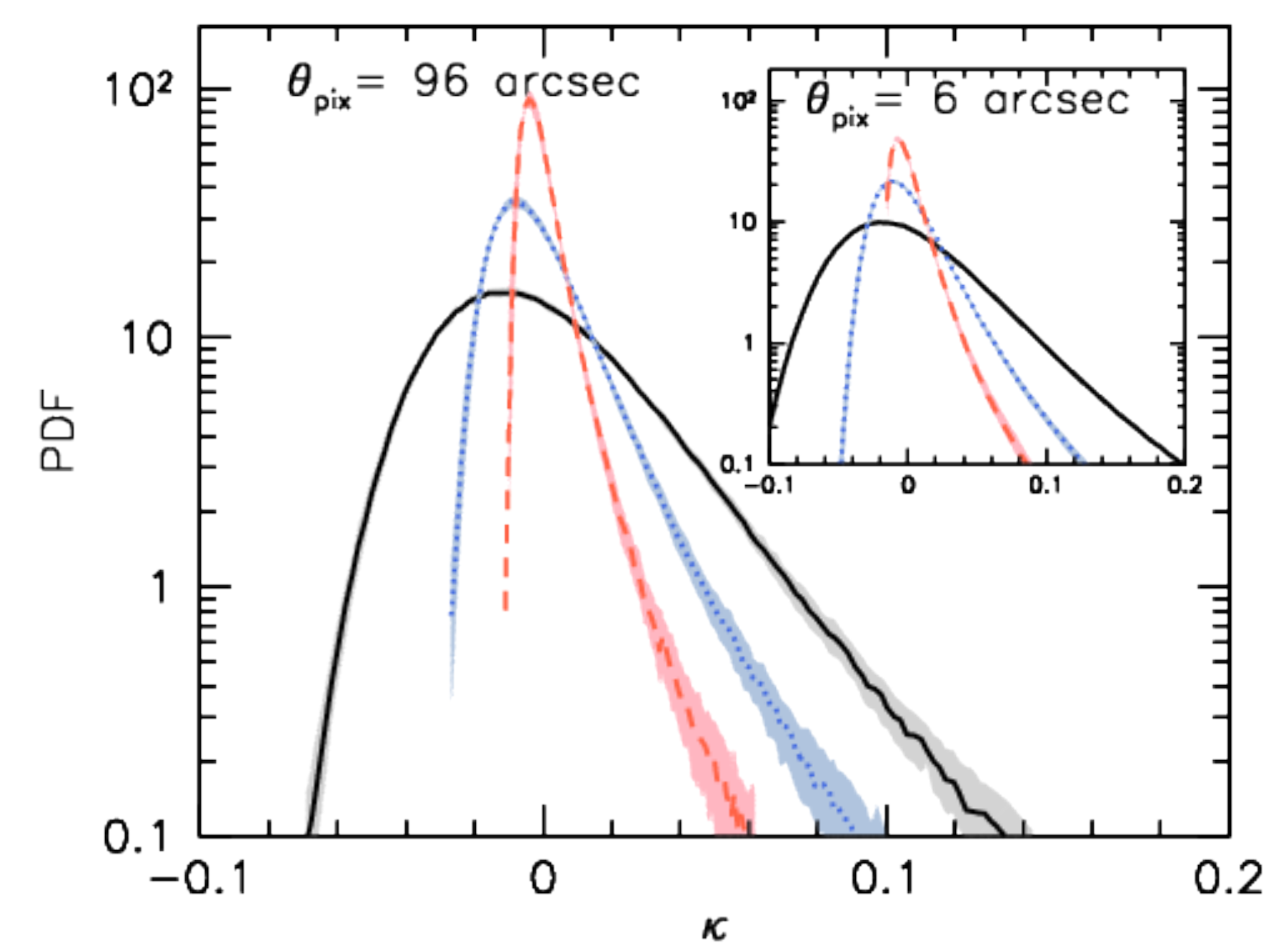}
\includegraphics[width=0.325\hsize]{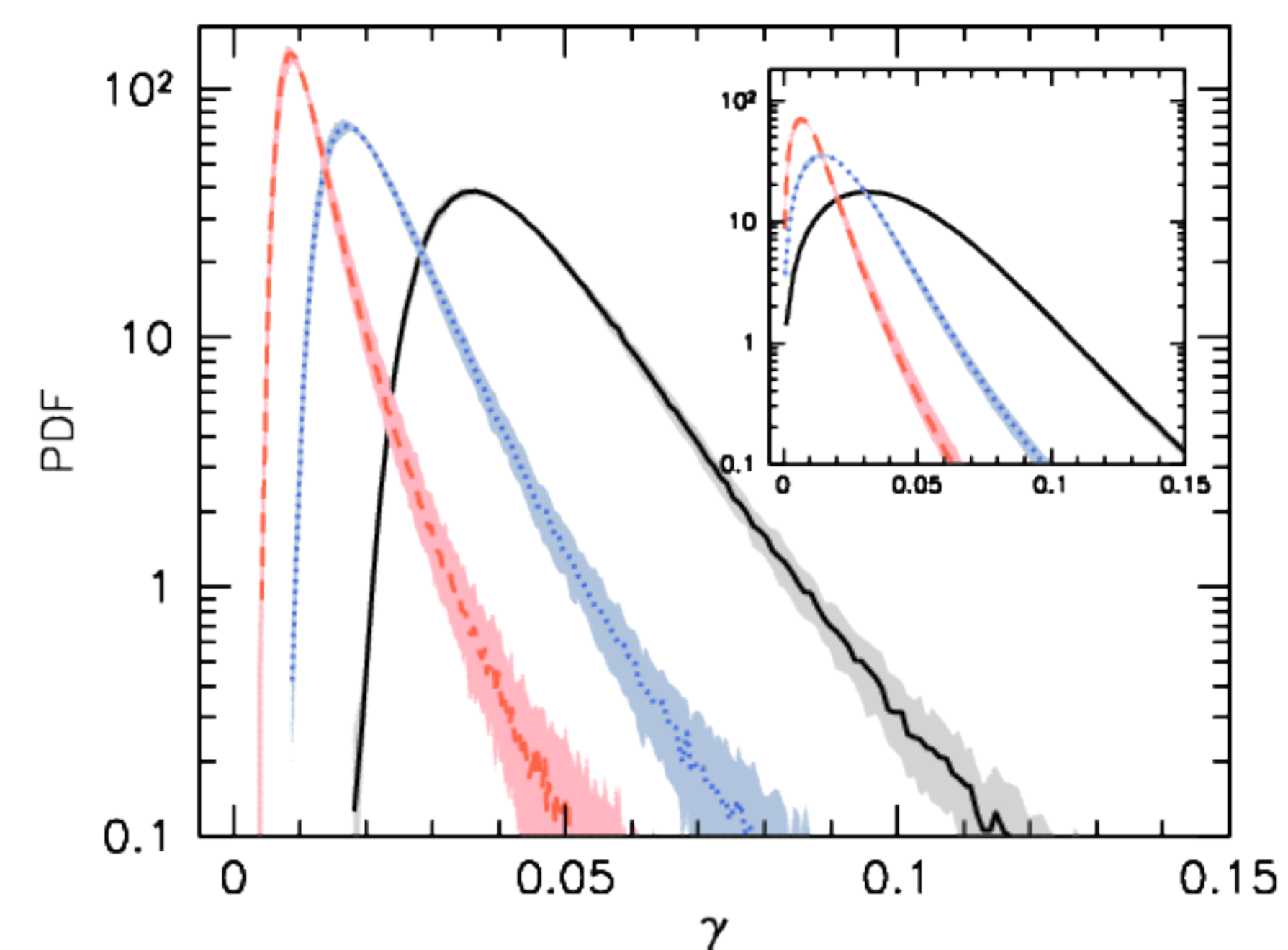}
\includegraphics[width=0.325\hsize]{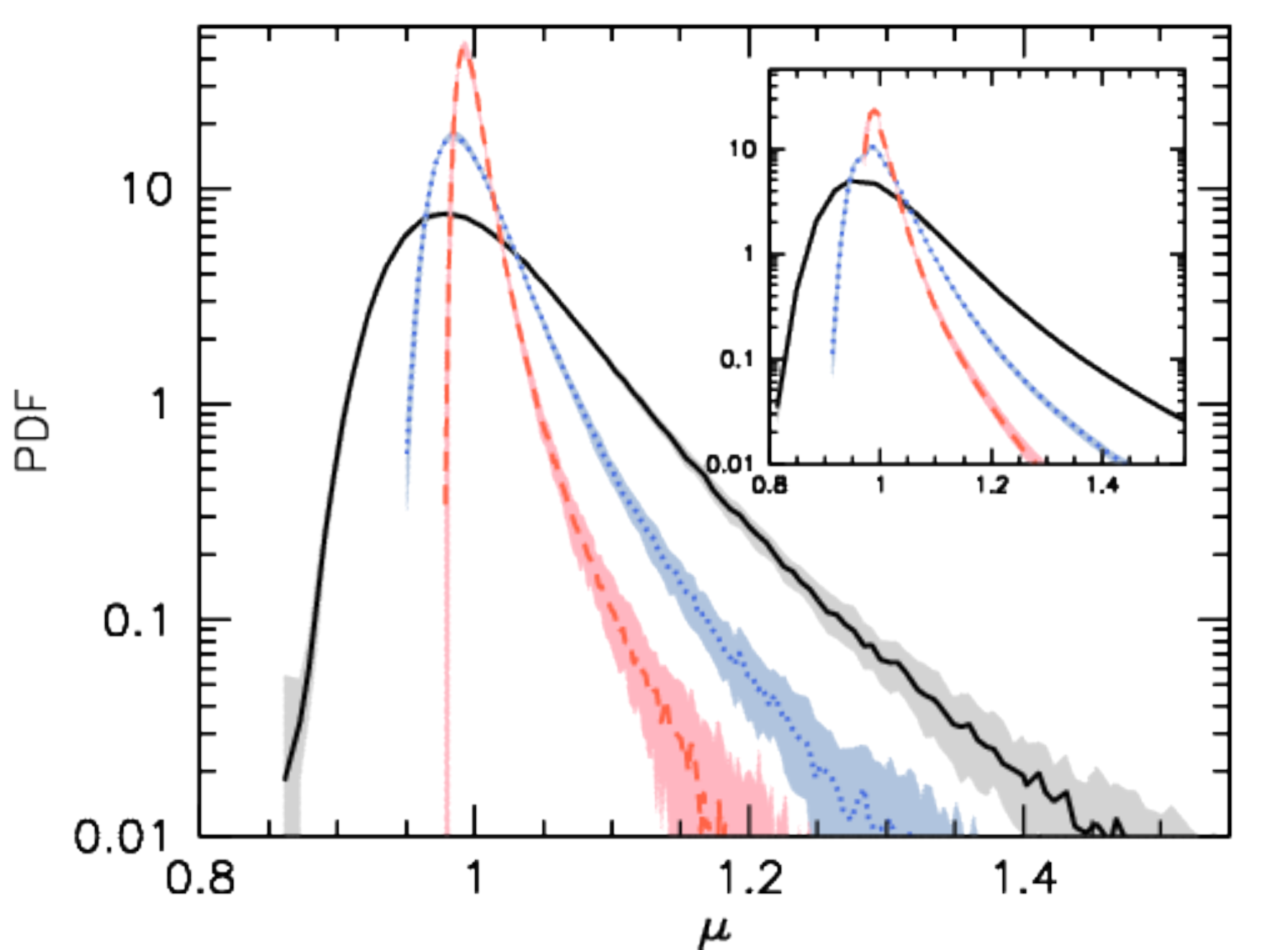}
\caption{Top  panels: median  convergence (left),  shear (centre)  and
  magnification (right) probability  distribution function measured in
  the different W1 and W4 light-cone realisations for sources at three
  different redshifts, from  the original maps with  resolution of $6$
  arcsec  per pixel.   The shaded  regions enclose  the first  and the
  second quartile of the distributions at fixed value.  \label{figpdf}
  Bottom  panels: median  distributions  obtained at  the three  fixed
  source redshifts  from the maps  having a resolution of  $96$ arcsec
  per pixel.  For  comparison, the small sub-panels show,  in the same
  axis  scales,  the  corresponding distributions  from  the  original
  resolution                                                    maps.}
\includegraphics[width=0.33\hsize]{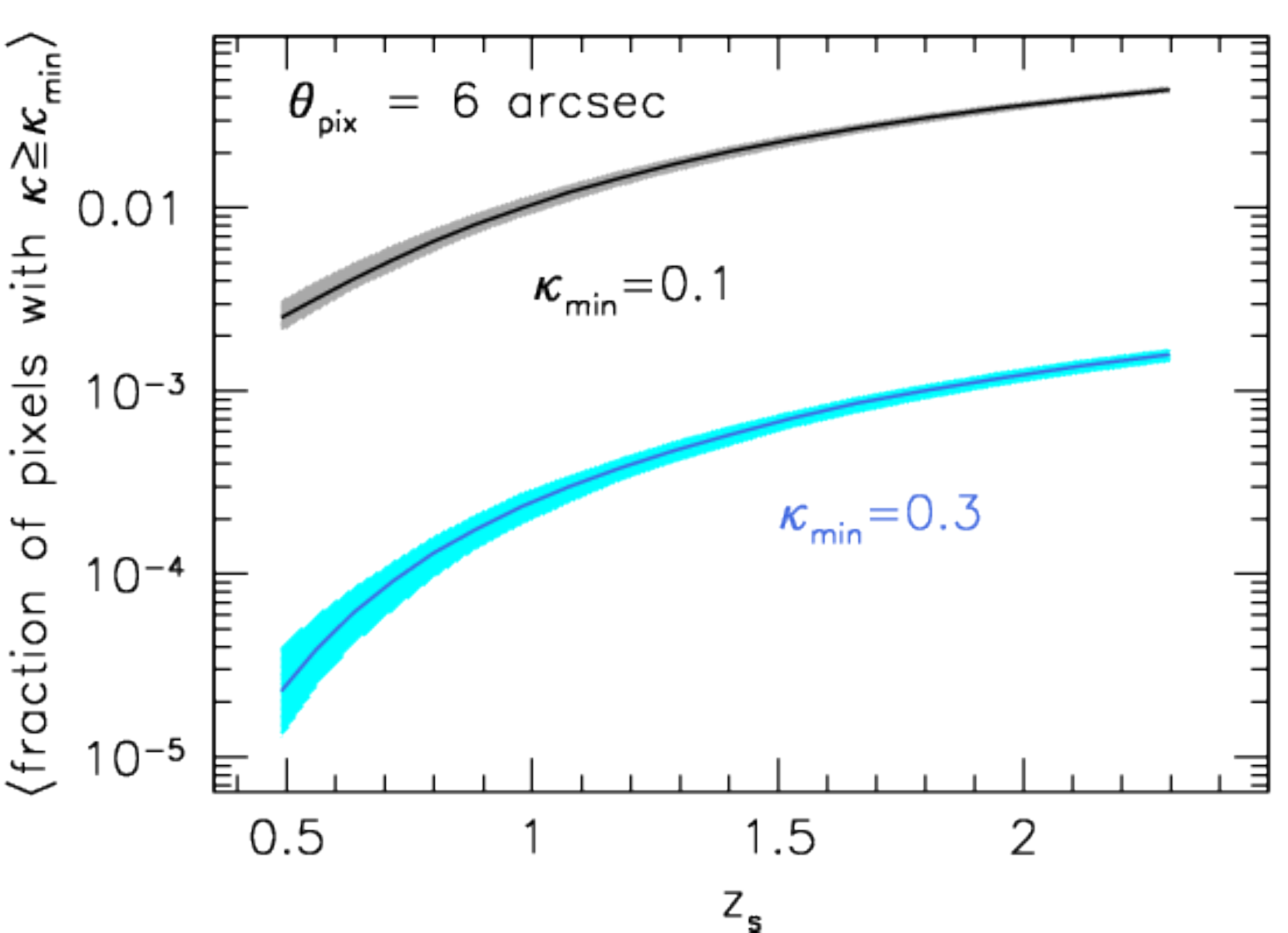}
\includegraphics[width=0.33\hsize]{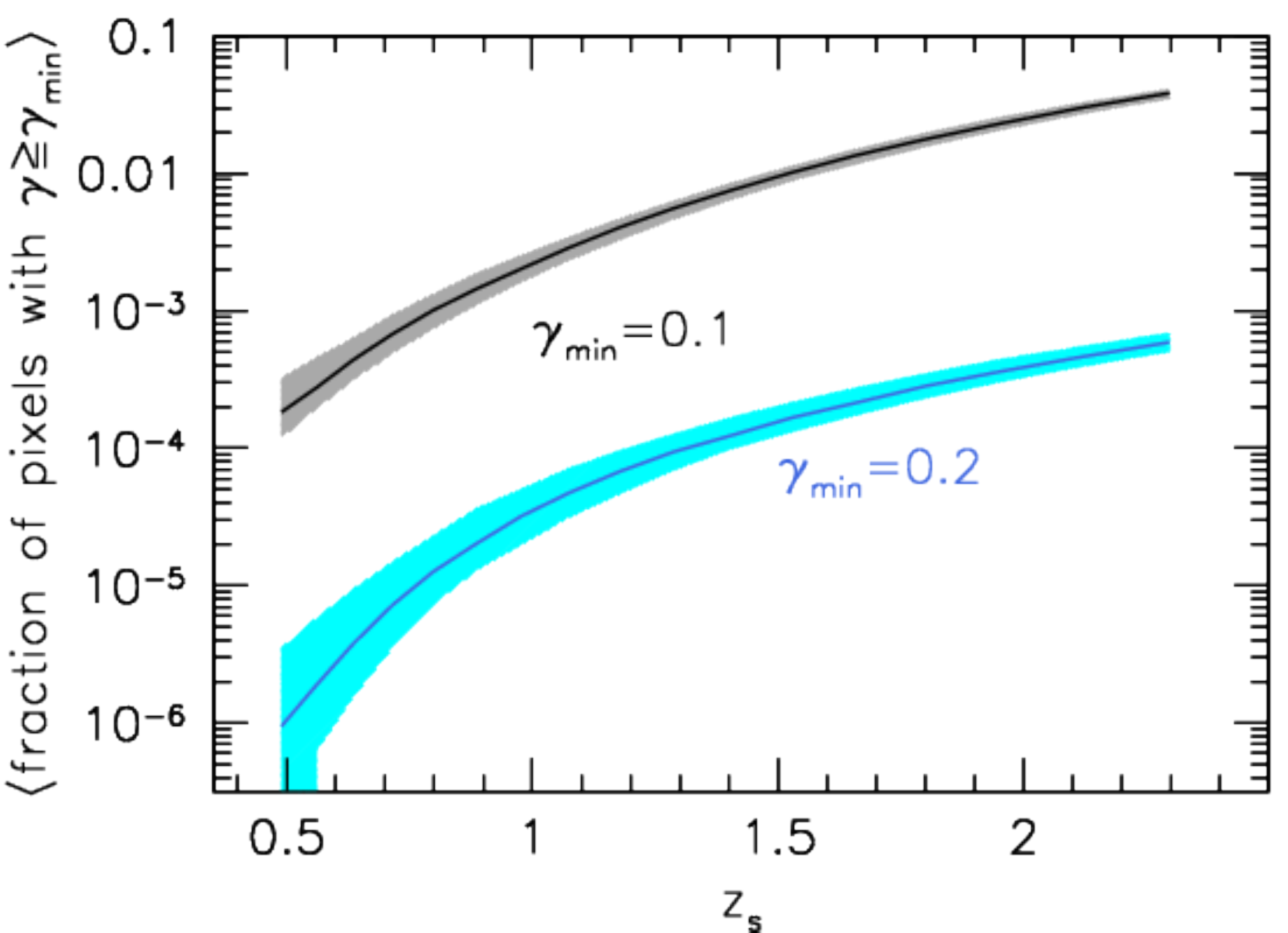}
\includegraphics[width=0.33\hsize]{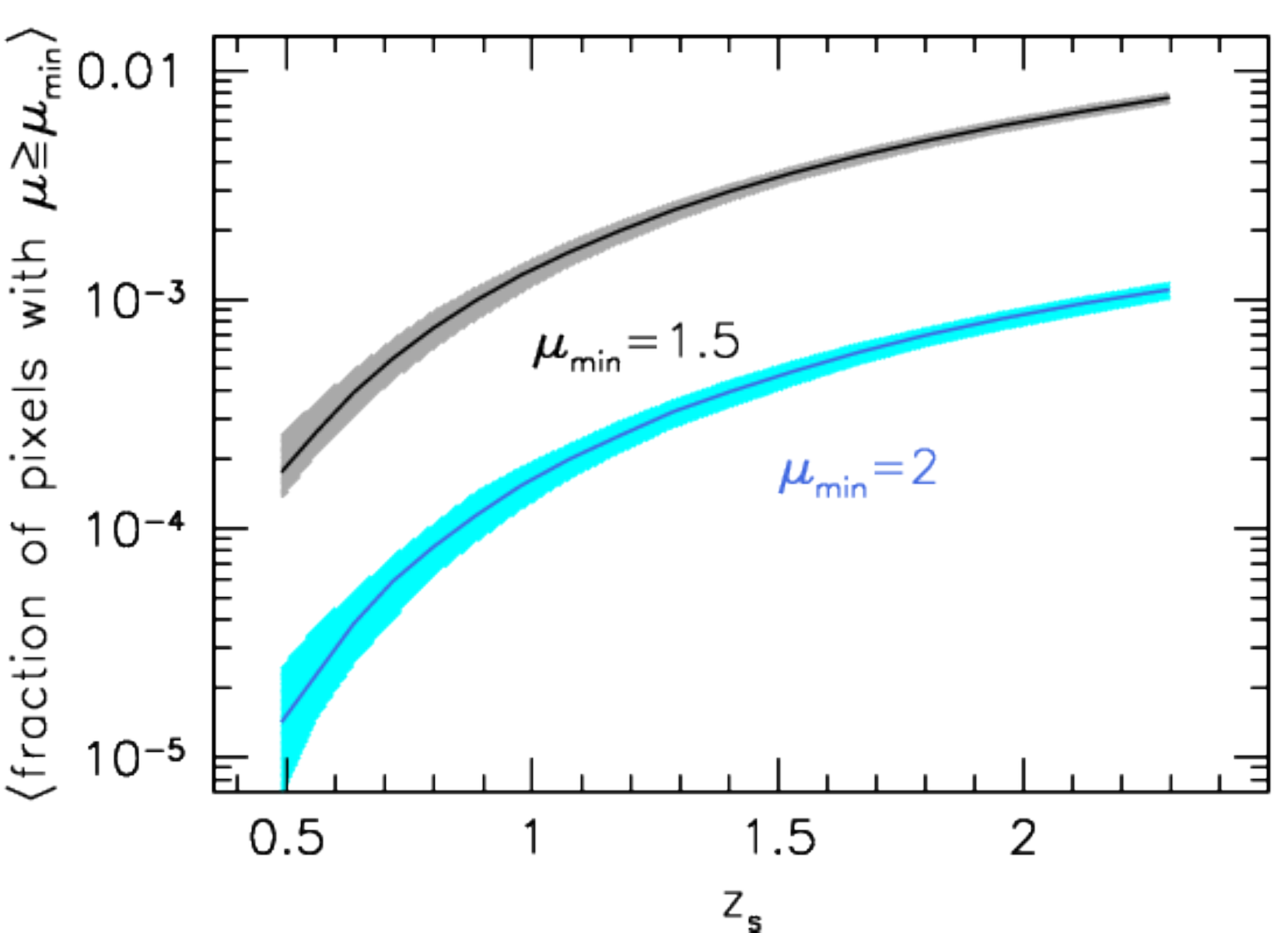}
\caption{Median fraction of pixels, among all the different light-cone
  realisations of W1  and W4, with convergence  (left), shear (centre)
  and  magnification (right)  above  a given  minimum  threshold as  a
  function of the source redshift. \label{figpdfthreshold}}
\end{figure*}

The one-point probability  distribution of the field can  also help in
distinguishing       the       underling      cosmological       model
\citep{takahashi11,pace15,giocoli15}.     In    the   top-panels    of
Fig.~\ref{figpdf} shows  the probability distribution function  of the
convergence (left), absolute shear  (centre) and magnification (right)
averaged over  the all  153 realisations  of the  two fields  from the
original maps having an pixel resolution of $6$ arcsec.

From  the  figure the  different  line  style curves,  with  different
corresponding colours, represent the  median over all the realisations
and the  shaded region encloses  the first  and the third  quartile at
fixed  values.  As  was  done  for the  power  spectrum,  we show  the
distributions for  three different  fixed source redshifts  $z=0.5$, 1
and 2.3. In the convergence PDF we notice that while the average value
of the field  remains null, the shape of the  distribution enlarges as
the source redshift increases giving also  rise to an high value tail.
The same  behaviour is  reflected in the  shear and  the magnification
even  if the  average value  of the  first increases  with $z_s$.   We
remind the reader that the distributions depends on the pixel sizes of
the maps  that have been constructed  which in this case  is 6 arcsec.
To  better interpret  those results,  in Fig~\ref{figpdfthreshold}  we
show  the median  fraction  of  pixels as  a  function  of the  source
redshift with convergence, shear  and magnification (left, central and
right  panel,  respectively) above  a  given  minimum threshold.   The
shaded  regions  enclose the  first  and  the  third quartile  of  the
distributions   at   fixed   $z_s$.    In   the   bottom   panels   of
Fig.~\ref{figpdf}  we  show the  one-point  distribution  of the  maps
degraded to a resolution of $96$  arcsec per pixels that, as discussed
already for the  cosmic shear power spectrum, corresponds  to scale at
which the  different source  redshift maps in  the light-cone  are not
affected  by  particle  shot-noise.   The small  sub-panels  show  the
one-point distributions  of the  maps with $\theta_{pix}=6$  arcsec in
the same axis scales, for a  more direct comparison.  From the figures
we  notice   that  the   distribution  function   of  the   maps  with
$\theta_{pix}=96$ arcsec are less spread  than those computed from the
maps with  $\theta_{pix}=6$ arcsec  for a  combination of  two effects
($i$) the quenching  of the particle noise mainly present  in the maps
with $z_s=0.5$ and for low values  of the distributions and ($ii$) the
disappearance of high density peaks due  to loss of the resolution and
of resolving cluster cores.

\begin{figure*}
\includegraphics[width=0.33\hsize]{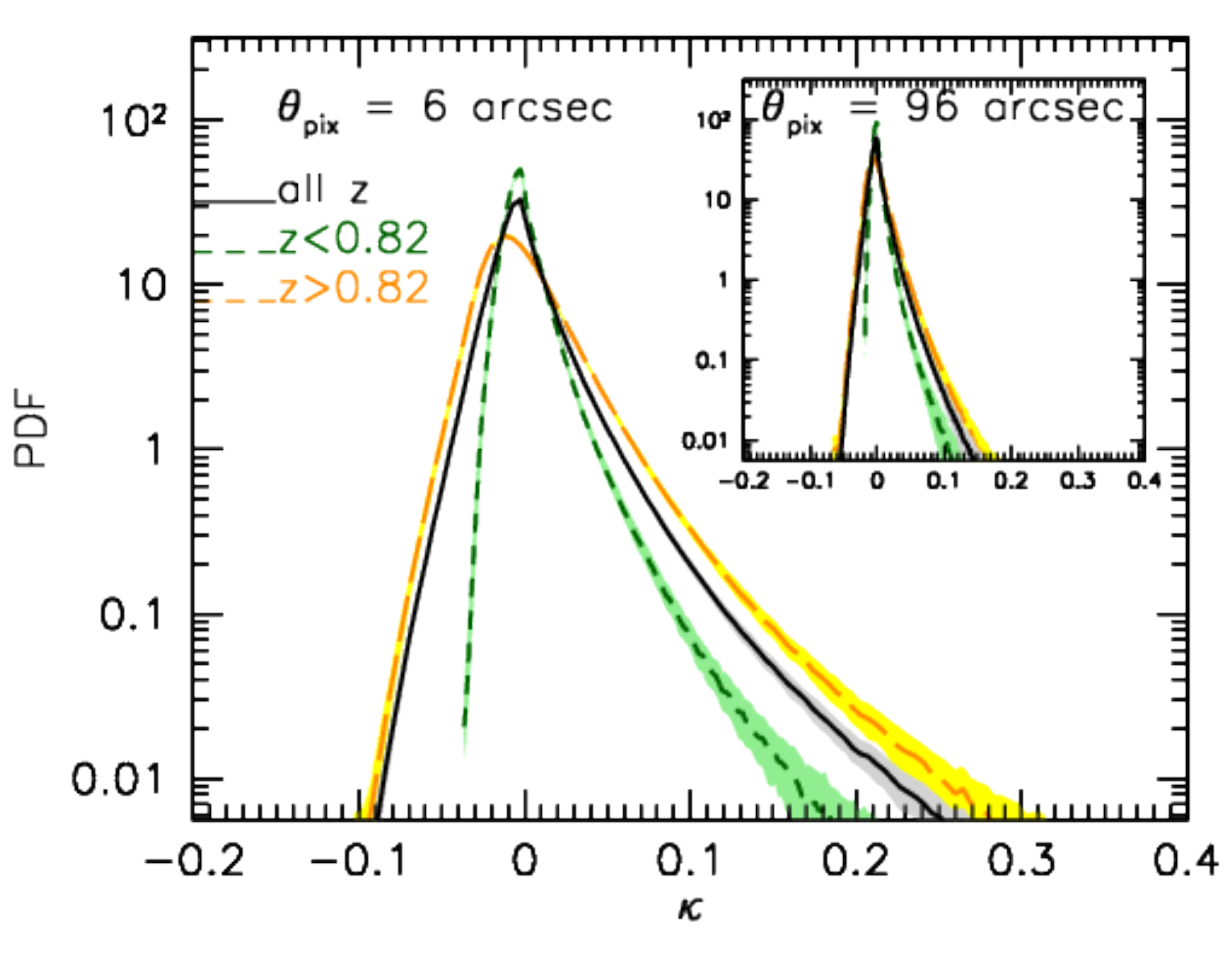}
\includegraphics[width=0.33\hsize]{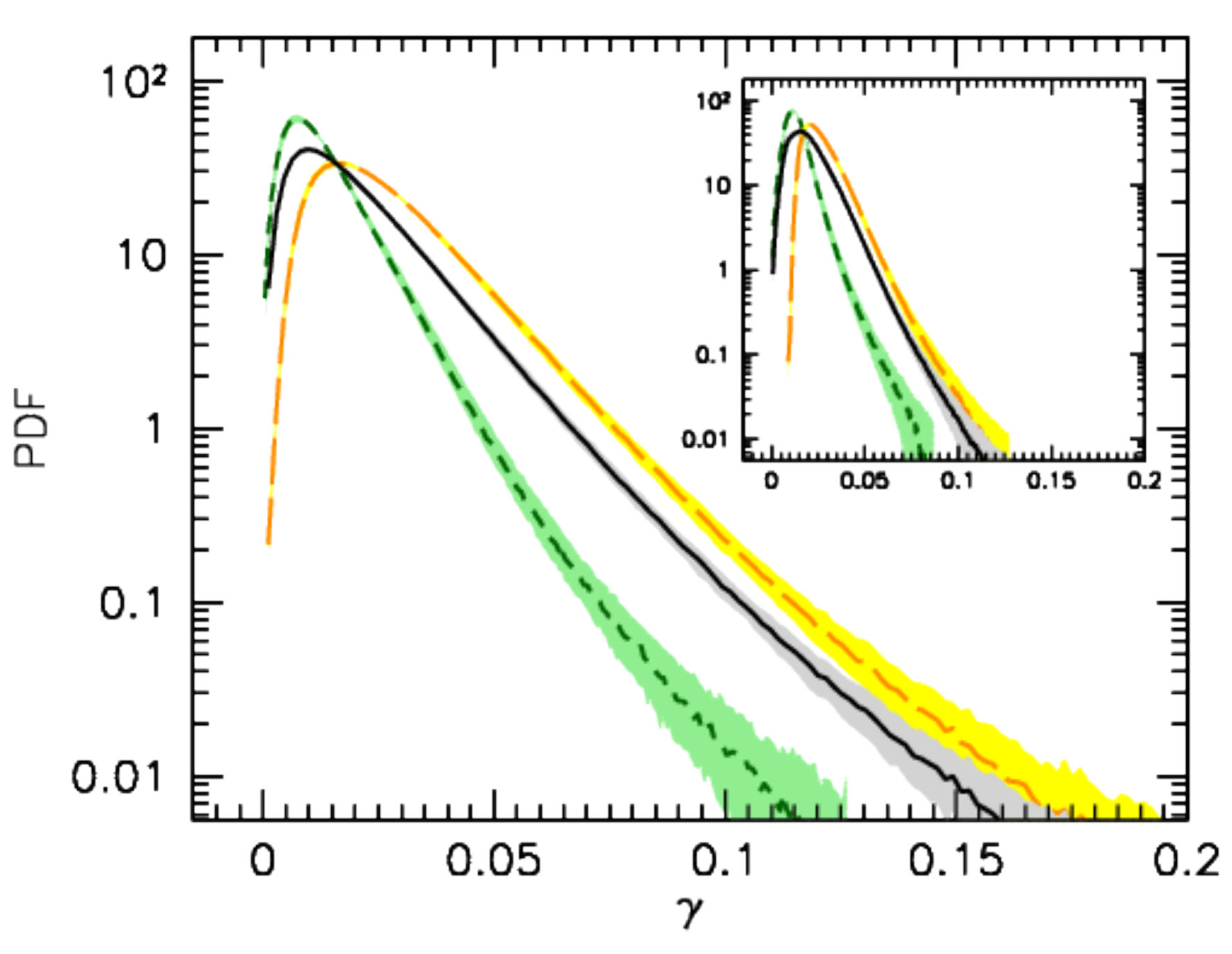}
\includegraphics[width=0.33\hsize]{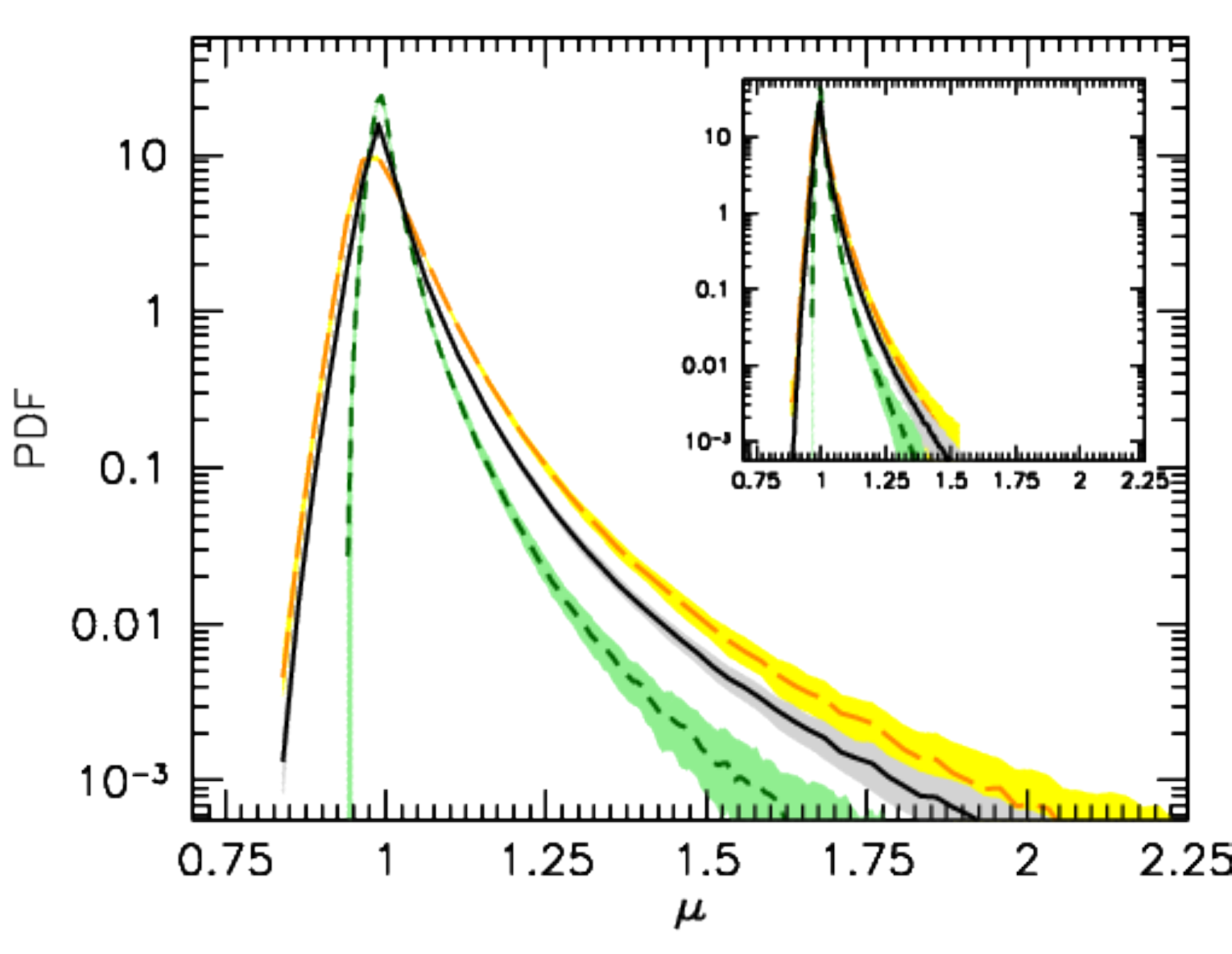}
\caption{Same as Fig.~\ref{figpdf} for maps extracted from the
  light-cones considering a CFHTLS-like source redshift distribution,
  considering the original maps with $\theta_{pix}=6$ arcsec.  While
  the black curves show the median of the total source sample, the
  long-dashed orange and the short-dashed green refer to sources above
  and below the median source redshift $\langle z \rangle$=0.87,
  respectively.  Shaded regions enclose the first and the third
  quartile of the distributions at fixed value.\label{fnzPDF} The
  small sub-panels show the same distributions from the maps with
  $\theta_{pix}=96$ arcsec, corresponding to scale at which the
  different source redshift maps in the light-cone are not affected by
  particle shot-noise.}
\end{figure*}

In  Fig.~\ref{fnzPDF}  we  show the  median  Probability  Distribution
Functions of convergence, shear  and magnification combining different
catalogues extracted from  the 54 and 99 realisations of  W1 and W4 --
using  the  original  maps  with  pixel  size  equal  to  $6$  arcsec,
respectively.   In particular  for each  field of  view we  generate 8
different  random  catalogues  sampling  the  CFHTLS  source  redshift
distribution and  randomly locating  the sources  within the  field of
view  for a  total of  1224  catalogues.  We  do not  account for  any
clustering in the source distribution since this is beyond the purpose
of this  first work.   The black  histograms display  the case  of the
whole source sample  up to redshift $z_s=2.3$, while  the short dashed
green  and the  long dashed  orange  refer to  the lensing  quantities
computed  for sources  below and  above the  median value  $z_m=0.87$,
respectively.   For each  histogram the  corresponding shaded  regions
enclose the first and the third quartile of the distributions at fixed
value.   The small  sub panels  show the  distributions from  the same
source  redshift distribution  obtained from  the maps  degraded by  a
factor of $16$ with respect to  the original ones.  As we have noticed
in  the  bottom   panels  of  Fig.~\ref{figpdf}  in   these  case  the
distributions  are less  spread and  miss of  the values  regime tails
\citep{takahashi11}.

\subsection{Other Statistics}

In this section, we study the predictions for other lensing statistics
from the simulated light-cones.  Besides the power spectrum, there are
other lensing  statistics that  can be  used to  probe both  the small
scale  matter density  distribution and  the dark  energy equation  of
state, as well as the  power spectrum normalisation.  Among these, the
top-hat  shear   dispersion  (equivalent   to  the  variance   of  the
convergence   field  convolved   with  a   top-hat  filter)   and  the
aperture-mass   dispersion  (equivalent   to  the   variance  of   the
convergence field  convolved with  a compensated aperture  filter) are
the most widely used used for cosmological investigation.  Comparisons
with  theoretical models  can be  easily done  considering that  these
quantities can be  analytically computed as weighted  integrals of the
convergence power spectrum.

The two filter that we will adopt to convolve the convergence maps are:
\begin{equation}
\tilde{W}_{TH} = \dfrac{2 J_1(l \theta)}{l \theta} 
\end{equation}
for the top-hat shear dispersion and 
\begin{equation}
\tilde{W}_{Ap} = \dfrac{\sqrt{276} J_4(l \theta)}{(l \theta)^2} 
\end{equation}
for the aperture-mass dispersion.  The functions $J_1(x)$ and $J_4(x)$
represent   the  first   and   the  forth   order  Bessel   functions,
respectively. To  clarify our  methodology, in the  the first  case we
will compute the second-order statistics from the following relation:
\begin{equation}
\langle \kappa^2 \rangle_{TH/Ap} = \dfrac{1}{2 \pi} \int \mathrm{d} l \,l P_{\kappa}(l) 
\tilde{W}^2_{TH/Ap}(l \theta)\, \label{eqk2};
\end{equation} 
considering the appropriate weight function in each case and using the $P_{\kappa}(l)$ computed
from the convergence maps.

\begin{figure*}
\includegraphics[width=0.45\hsize]{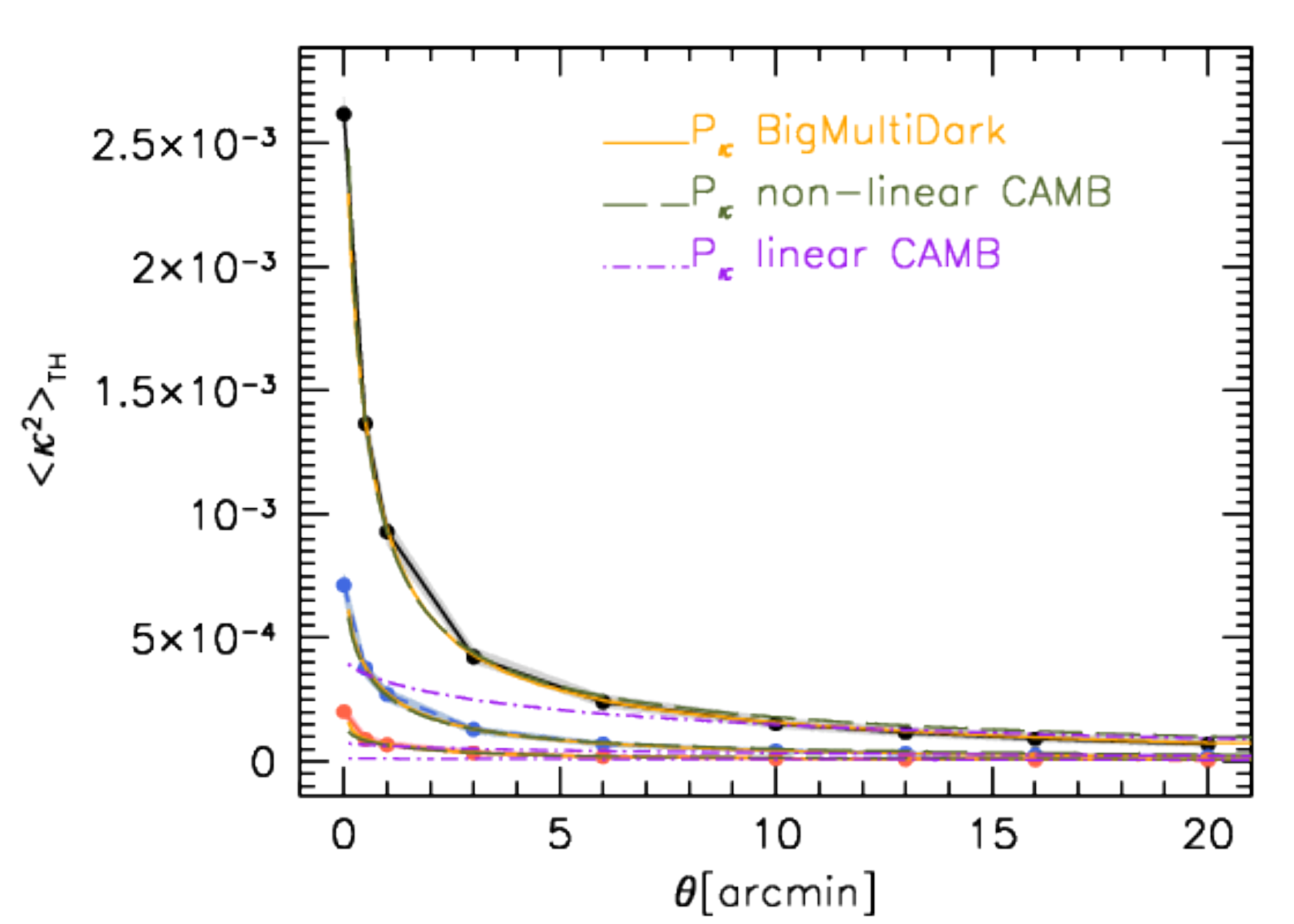}
\includegraphics[width=0.45\hsize]{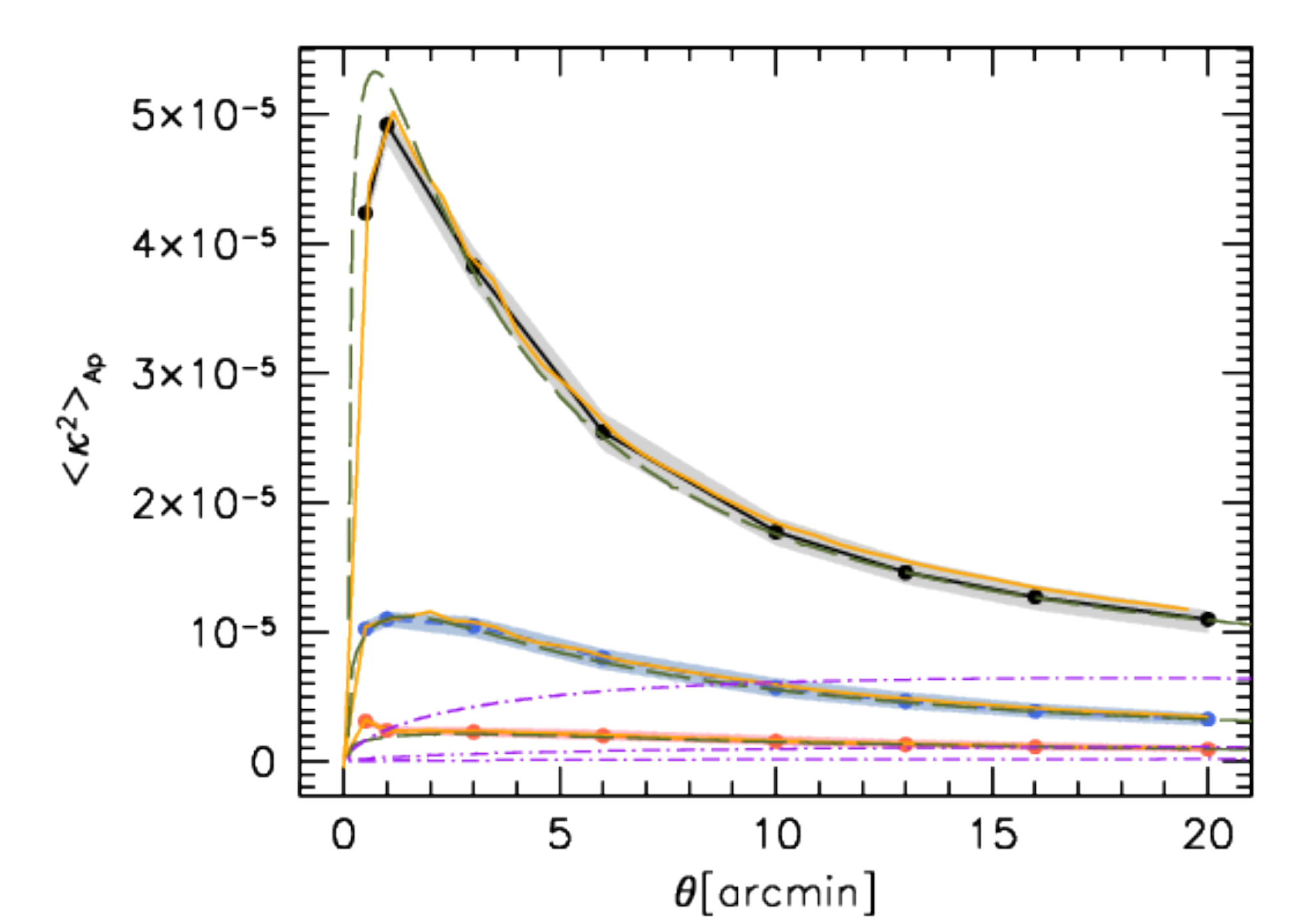}
\caption{Variance of the convergence field  in a top-hat (left) and in
  a  compensated  aperture (right)  computed  averaging  over all  the
  light-cone  realisations.  The  various line  styles refer  to three
  different  source redshifts  considered $z=2.3$,  1 and  0.5: solid,
  dashed and long-dashed, respectively.   The shaded regions represent
  the variance  of the measurements  in all realisations.   The orange
  solid  curves show  the  predictions computed  from the  convergence
  power spectrum at the corresponding redshifts. \label{k2tha}}
\end{figure*}
These statistics  can be calculated  by either applying the  filter to
the two dimensional convergence field and then finding the variance or
by  finding  the power  spectrum  of  the  field  and then  doing  the
integrals above, the latter being much  faster on large fields.  It is
possible that numerical effects could  make these results different in
practice.  As a  cross-check we calculated them in both  ways and find
they are in very good agreement as will be seen in the plots.

In  Fig.~\ref{k2tha}  we  show  the   top-hat  shear  (left)  and  the
aperture-mass  dispersion   (right)  as  measured  in   the  different
realisations  of  the  two  considered  light-cones  with  sources  at
redshift $z_s=0.5$,  1 and 2.3  -- the corresponding  redshift colours
have  been  chosen  as   in  Fig.~\ref{figpdf}.   The  coloured  lines
connecting the data points show the median measurements from the maps,
at each corresponding redshift,  with the corresponding shaded regions
enclosing  the  first   and  the  third  quartile   on  the  different
realisations  --   slow  computation.   We  remind   the  reader  that
aperture-mass dispersion data for  $z_s=0.5$ and $\theta \lesssim 1.6$
arcmin may  be affected  by particle shot-noise.  From the  other side
small scale particle  shot-noise does not appear  for the measurements
for $z_s=1$ and $2.3$.

The solid  orange curves, instead,
display the measurements done  adopting the computed convergence power
spectra from the  maps in equation (\ref{eqk2}). The  dashed green and
the dot-dashed  magenta curves show  the predictions --  from equation
(\ref{eqk2})  --  adopting  linear  and non-linear  power  spectra  as
implemented  in \textsc{CAMB}.   Apart for  the linear  case, all  the
measurements are in  quite good agreement with  each other, indicating
the  importance of  non-linear  modelling on  the  small scale  lensing
measurements.  In the case of a top-hat filter the differences between
the linear  and nonlinear  results start to  be significant  on scales
below $\sim 5-7$ arcmin -- depending  on the source redshifts, for the
compensated aperture the deviations are  already evident at 20 arcmin.
These  behaviours  are  related  to the  relative  weighs  assigned  to
different scales  by the  two considered  filters. In  particular, the
compensated   aperture   filter   is   very   peaked   enhancing   the
non-linearities of the matter  density distribution at larger $\theta$
\citep{schneider98}.
\begin{figure*}
\includegraphics[width=0.45\hsize]{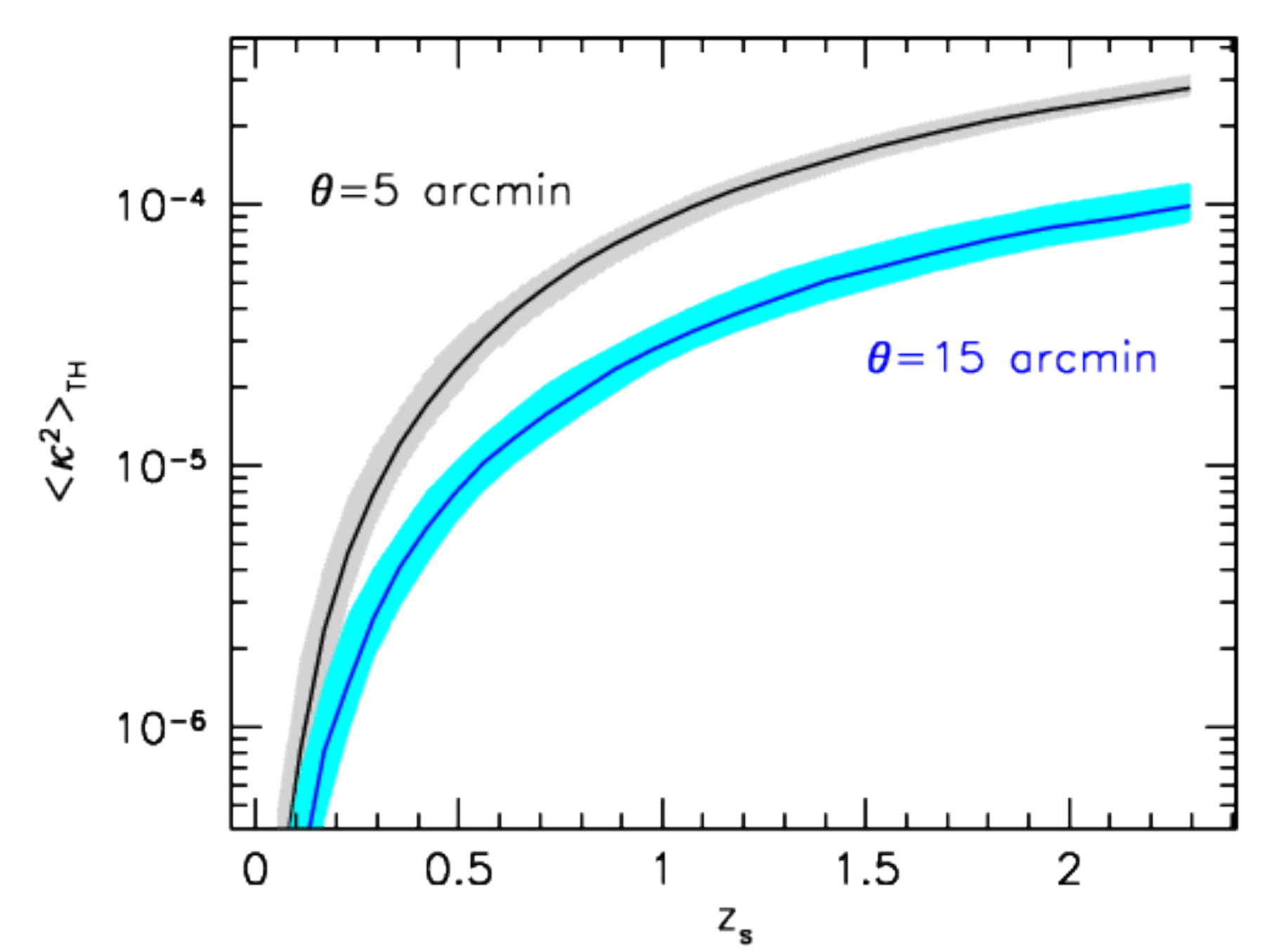}
\includegraphics[width=0.45\hsize]{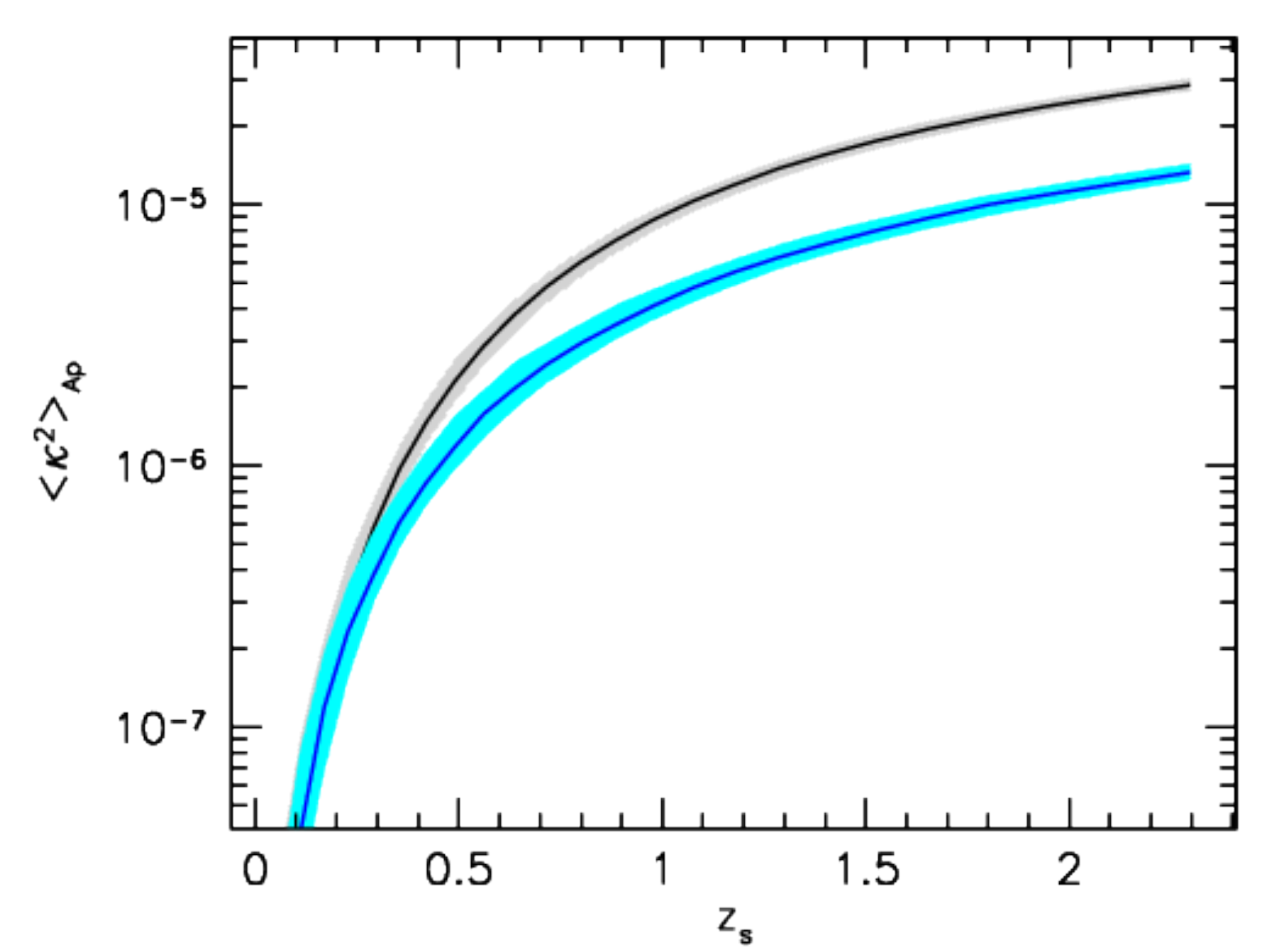}
\caption{Median  variance in  a top-hat  (left) and  in a  compensated
  aperture (right) of the convergence maps as a function of the source
  redshifts for  two different  smoothing scales: $\theta=5$  and $15$
  arcmin.  The curves  represent  the median  of  the measurements  as
  performed from the  different simulated lensing maps  and the shaded
  regions enclose the first and the third quartile of the distribution
  at fixed source redshift. \label{figk2thaz}}
\end{figure*}

Despite  the   close  agreement   between  analytic  theory   and  the
simulations  there  are   differences  which  can  be   seen  in  Fig.
~\ref{k2tha}.  They are  most pronounced in the right  hand panel, for
the aperture-mass dispersion, at small smoothing scale filters and for
$z_s=2.3$.  This discrepancy was also  noticed at high redshift in the
3D matter power spectrum as presented in Fig.~\ref{figpk}.

For sources at higher redshift the fluctuations in the convergence are
larger.  As shown  in Fig.~\ref{figk2thaz} both the  top-hat shear rms
and the aperture mass dispersion grow by about two orders of magnitude
between a source  redshift of $z_s=0.2$ and $z_s=2$. In  the figure we
present the median measurement for two filtering scales $\theta=5$ and
15 arcmin  -- representing  the typical  scales enclosing  the central
region of a galaxy cluster; where  we notice that for smaller $\theta$
the difference  of the non-linear  contributions between low  and high
redshifts structures is more enhanced than for larger filters.

\begin{figure*}
\includegraphics[width=0.45\hsize]{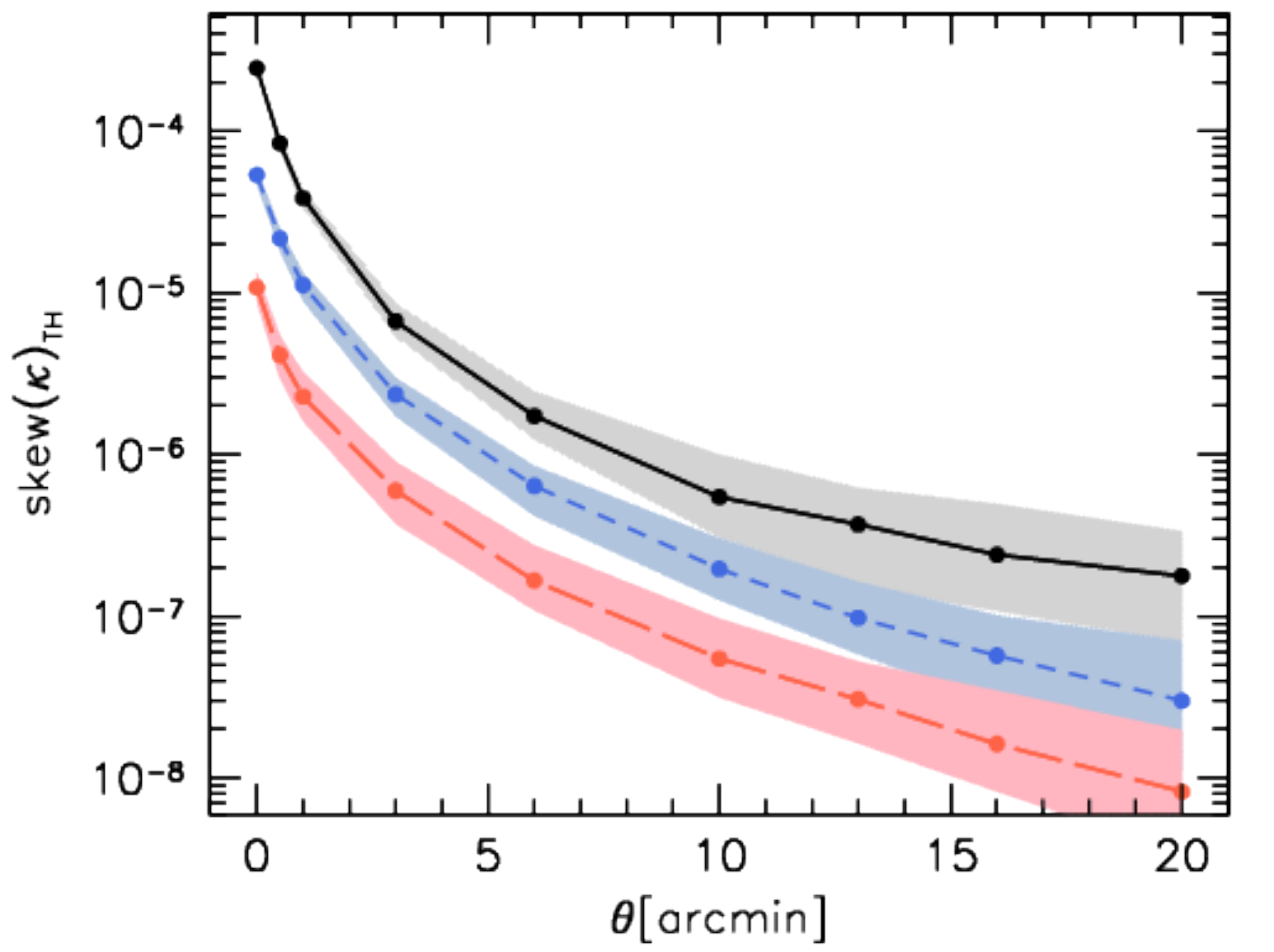}
\includegraphics[width=0.45\hsize]{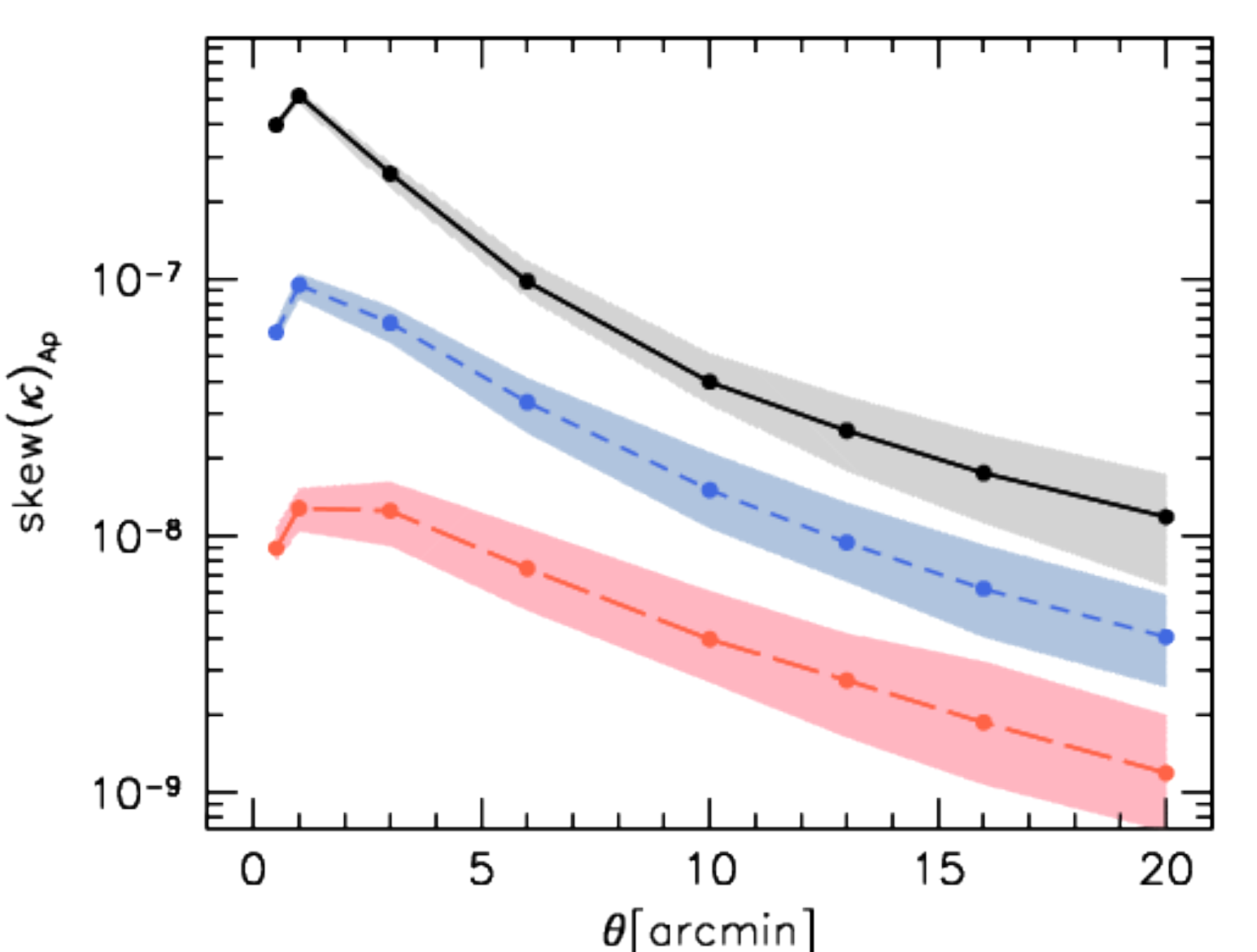} 
\caption{Skewness of the filtered  convergence maps with a
    top-hat (left) and  an aperture (right) function  measured on all
  the light-cone  realisations as  a function  of the  smoothing scale
  $\theta$.  The different line styles refer to three different source
  redshifts  considered   $z=2.3$,  1  and  0.5:   solid,  dashed  and
  long-dashed, respectively.\label{k3tha}}
\end{figure*}

\begin{figure*}
\includegraphics[width=0.45\hsize]{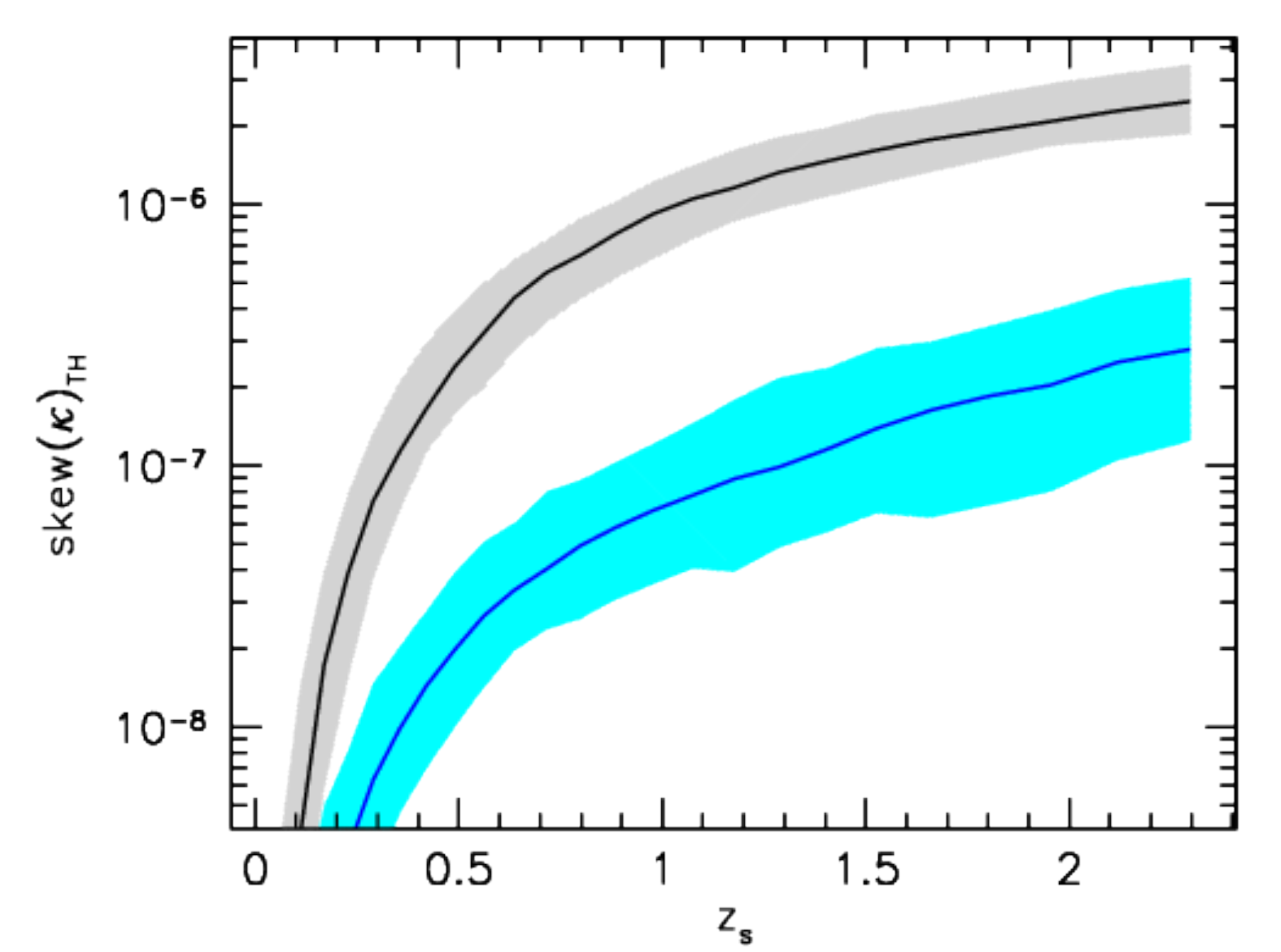}
\includegraphics[width=0.45\hsize]{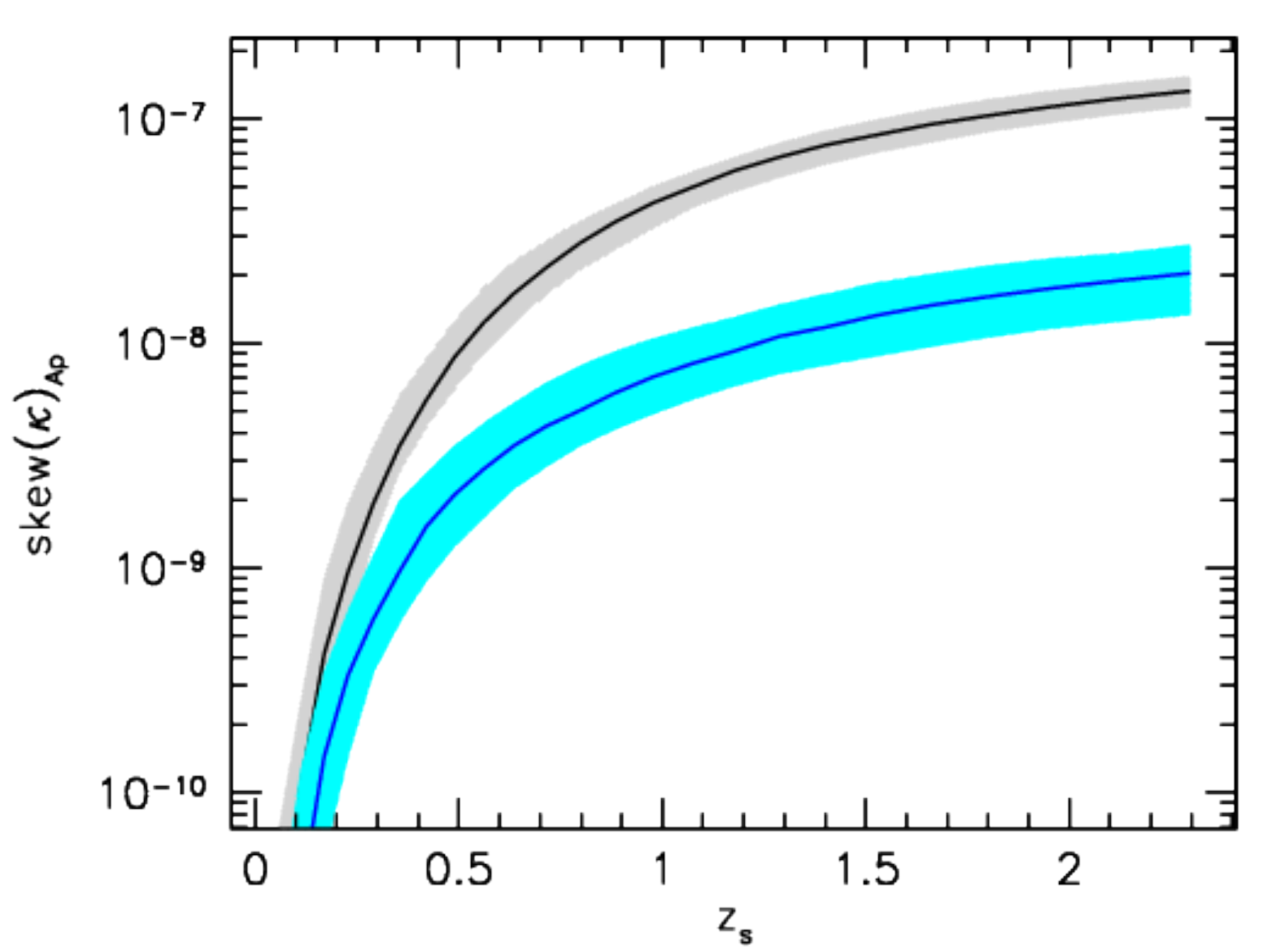}
\caption{Median skewness in  a top-hat (left)  and in a  compensated aperture
  (right)  of  the  convergence  maps  as a  function  of  the  source
  redshifts for  two different  smoothing scales: $\theta=5$  (higher) and $15$
  arcmin (lower). \label{figk3thaz}}
\end{figure*}

Another important  probe for understanding the  small scale clustering
of dark  matter and evolution of  dark energy evolution are  the three
point statistics.  We compute the skewness of the convergence maps for
each light cone  realisation both for W1 and W4,  for different source
redshifts and filtering scales $\theta$ computing:
\begin{equation}
 \mathrm{skew}(\kappa)_{TH/Ap} = \dfrac{\langle \kappa^3 \rangle_{TH/Ap}} 
{\langle \kappa^2 \rangle^{3/2}_{TH/Ap}}
 =  \dfrac{\dfrac{1}{n-1} \sum_i \left( \tilde{\kappa}_i - \langle \tilde{\kappa} 
 \rangle \right)^3 }{\left[\dfrac{1}{n-1}\sum_i  \left( \tilde{\kappa}_i - \langle \tilde{\kappa} \rangle \right)^2 \right]^{3/2}}
\end{equation}
where the  sum is extended  up to $n$, equals  to the total  number of
pixels in  each map  and $\tilde{k}$  represent the  filtered converge
map.   As  done  previously,  we   adopt  both  the  top-hat  and  the
compensated aperture  filter.  In Fig.~\ref{k3tha} we  show the median
of  the skewness  for three  different source  redshifts both  for the
top-hat (left)  and the  compensated aperture filter  (right).  Black,
blue and red  show the cases for sources at  redshift $z_s=2.3$, 1 and
$0.5$, with the  corresponding shaded regions enclosing  the first and
the third quartile of the  distribution. The redshift evolution of the
skewness, for the two filters,  is shown in Fig.~\ref{figk3thaz} again
for $\theta=5$ (black) and $15$ arcmin (blue).

\begin{figure}
\includegraphics[width=\hsize]{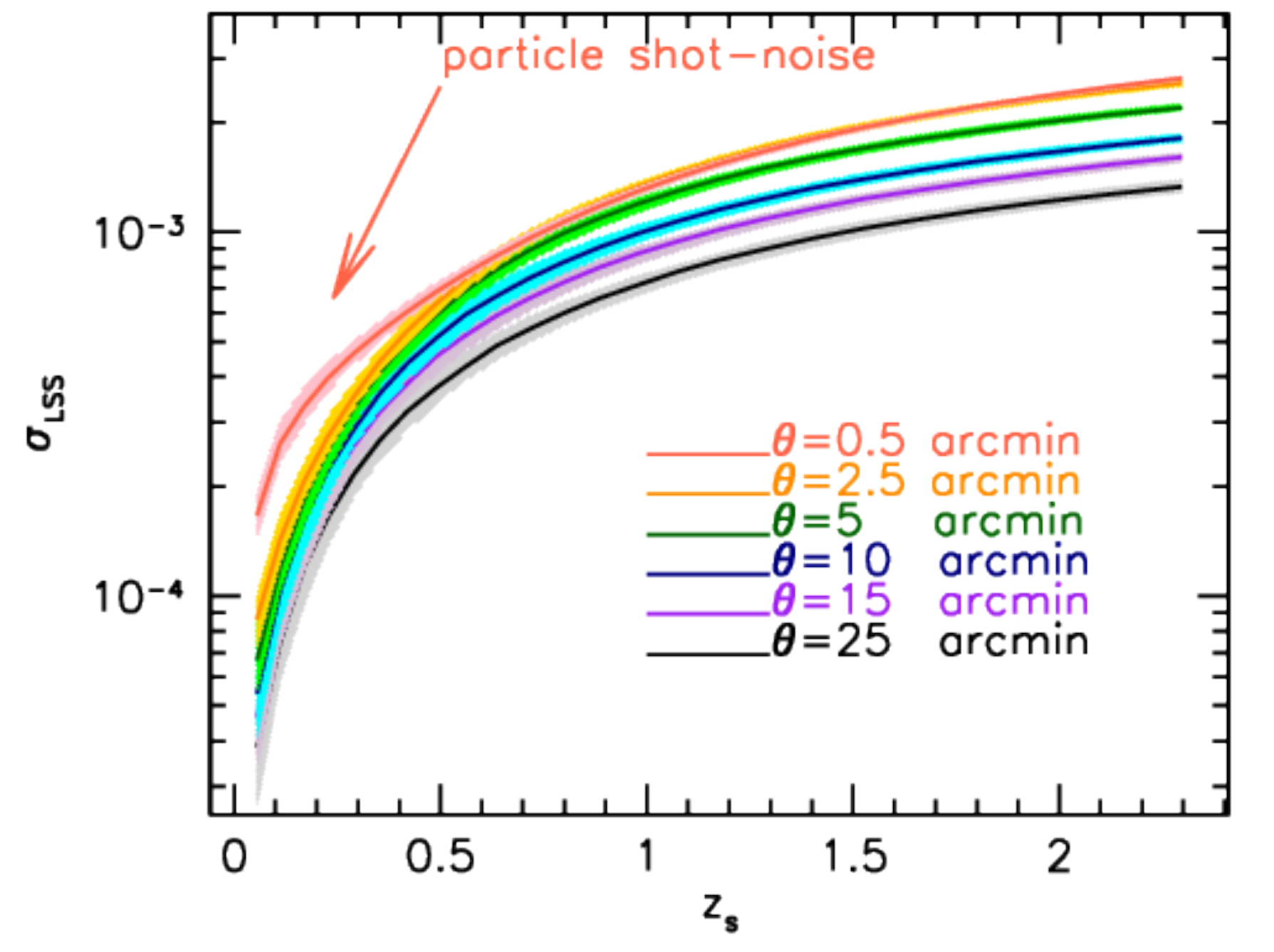}
\caption{Large  scale  structure noise  in  the spherical averaged 
shear profile as a function of the source redshift and for different filtering scales.
  \label{figsigmalss}}
\end{figure}

The last statistic  that we address in this section  regards the noise
from large scale structures to the  average tangential shear in a thin
circular annuls.  This quantity is  important to understand  the noise
induced  by  uncorrelated  structure   along  the  line-of-sight  when
measuring the  tangential shear  profile of a  galaxy cluster  and the
accuracy  with  which its  mass  and  concentration can  be  recovered
\citep{hoekstra11,giocoli14,petkova14}.  Typically  this noise depends
on the  scale we are looking  at and so  tends to be different  in the
weak and  the strong lensing regions  of a galaxy cluster.   While the
strong lensing signal  appears toward the core of  the galaxy cluster,
typically  scales well  below an  arcminute, the  weak lensing  signal
extends out  to larger scales --  from some arcminutes and  above.  In
order to  compute this noise  from our simulated light-cones  we adopt
the formalism as presented by \citet{hoekstra03} where the convergence
field is  convolved with the  following filtering function  in Fourier
space:
\begin{equation} \label{eq:cl_mass_error}
 g(l \theta) = \dfrac{J_2(l\theta)}{2 \pi}\,,
\end{equation}
where $J_2$  represents the second  order Bessel function.   The noise
from Large Scale Structures (LSS)  $\sigma_{LSS}$ will be given by the
square root  of the variance  of the convergence field  convolved with
$g(l\theta)$.  We  do this for each  realisation of the W1  and the W4
fields     and     each     considered    source     redshift     (see
Fig.~\ref{figplanes}). In Fig.~\ref{figsigmalss} we present the median
noise  by LSS  as a  function of  the source  redshifts for  different
filtering scales $\theta$.  At fixed redshift the noise decreases as a
function of  $\theta$ and  it increases  as a  function of  the source
redshift for  fixed $\theta$.  In the  figure we notice also  that the
measurements for $\theta=0.5$ arcmin at  for $z_s<0.75$ present a hump
due to  the discreetness of  the particle density distribution  in the
simulation.
From the figure we can see that from low ($z_s$=0.2) to high ($z_s=2$)
redshift sources  the noise  in the  spherical averaged  shear profile
increases by approximately one orders of magnitudes.

\subsection{Halo-shear lensing signal}

\begin{figure*}
\begin{tabular}{cc}
\includegraphics[width=0.45\hsize]{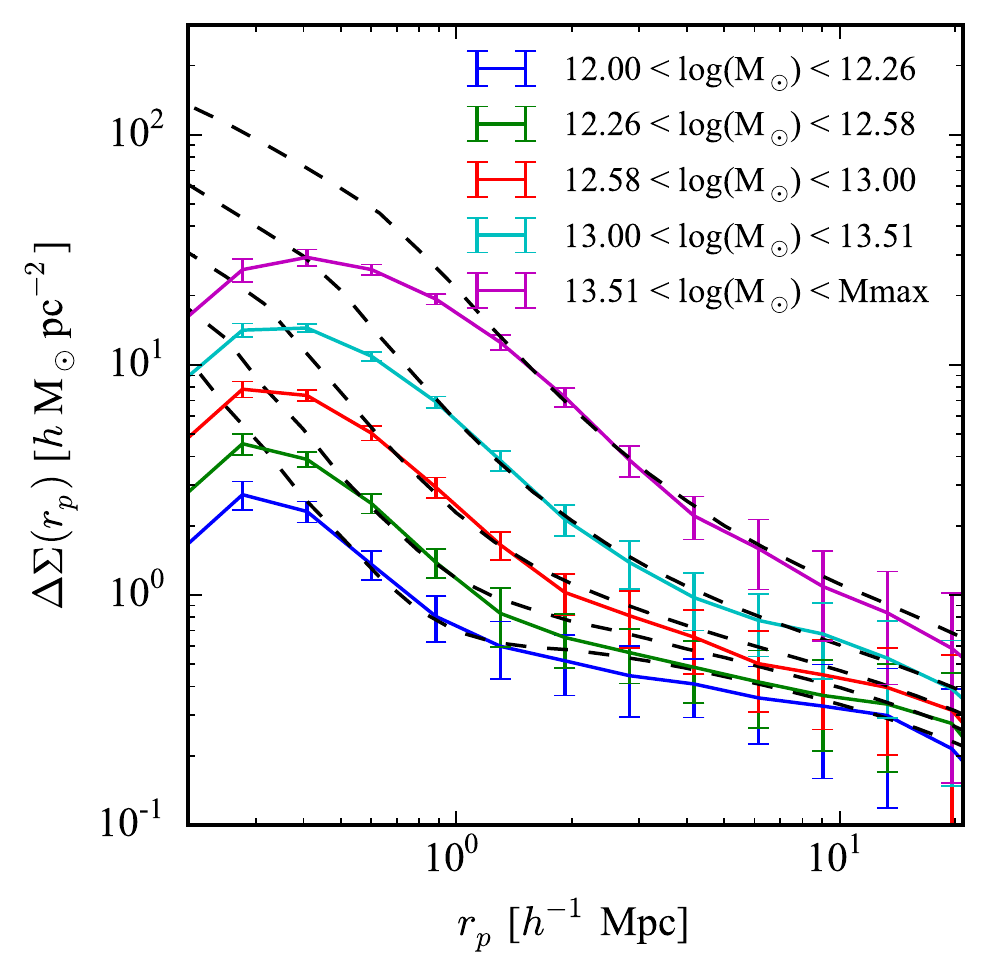}&
\includegraphics[width=0.45\hsize]{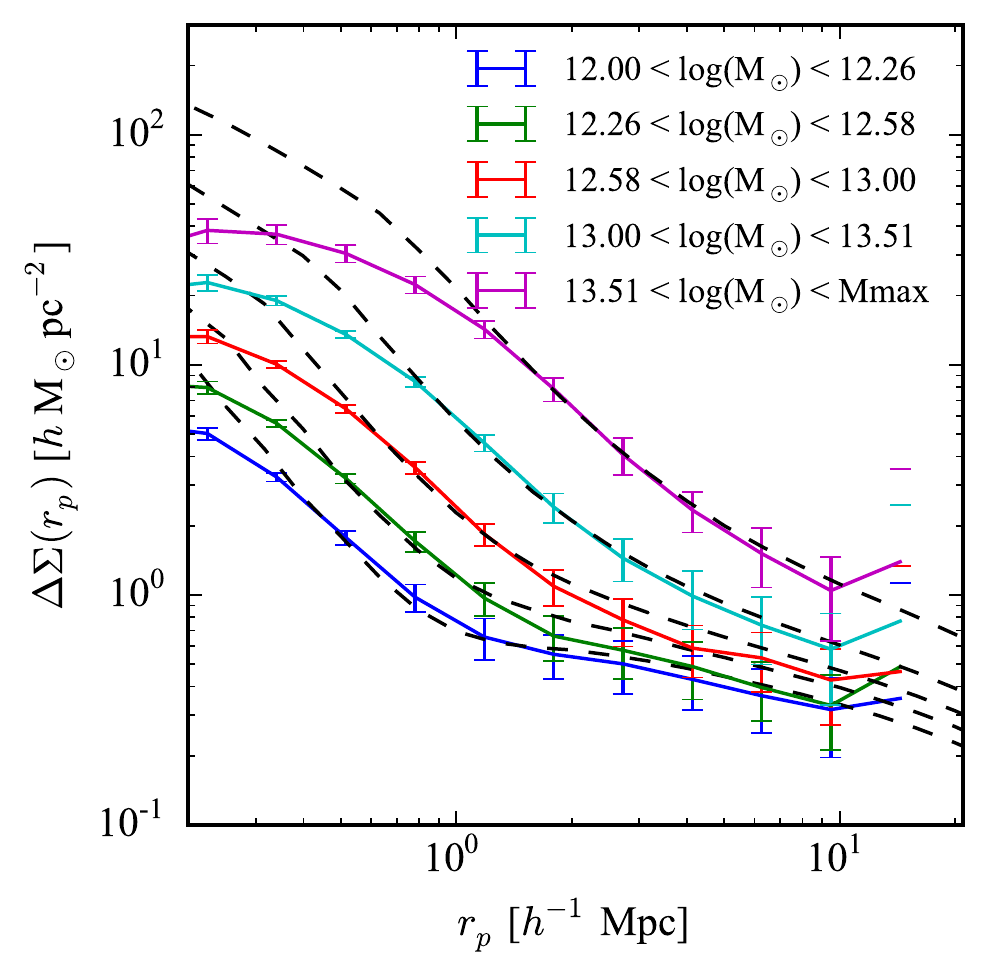}
\end{tabular}
\caption{{\it (left)} Halo-galaxy lensing signal measured for halos in
  different mass  bins at redshift  $z=0.52$ in field  W1. Theoretical
  predictions  are  shown in  dashed  lines.  {\it (right)}  Projected
  density  contrast profile  obtained  by stacking  the density  field
  around  halos of  the same  mass  bin. Distances  and densities  are
  computed in  comoving coordinates  to ease  the comparison  with the
  halo model predictions. \label{fig:gglensing}}
\end{figure*}

\begin{figure*}

\begin{tabular}{cc}
\includegraphics[width=0.45\hsize]{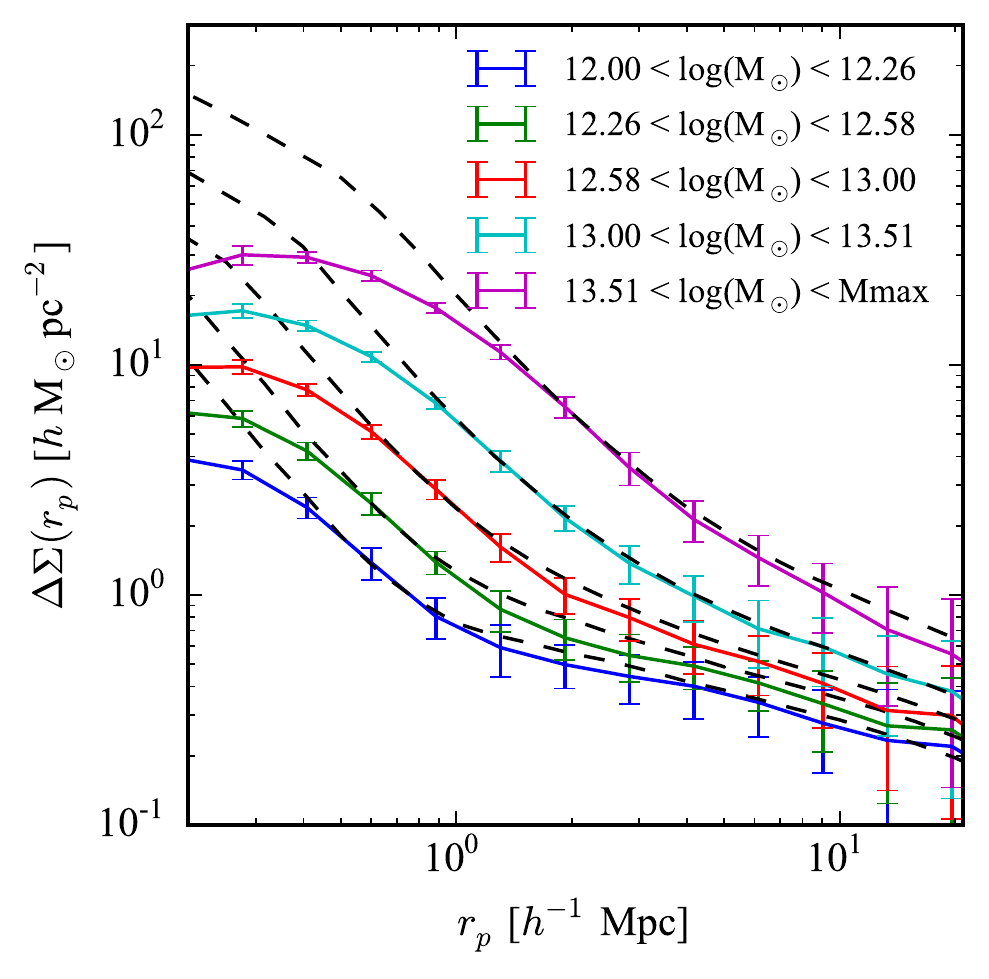}&
\includegraphics[width=0.45\hsize]{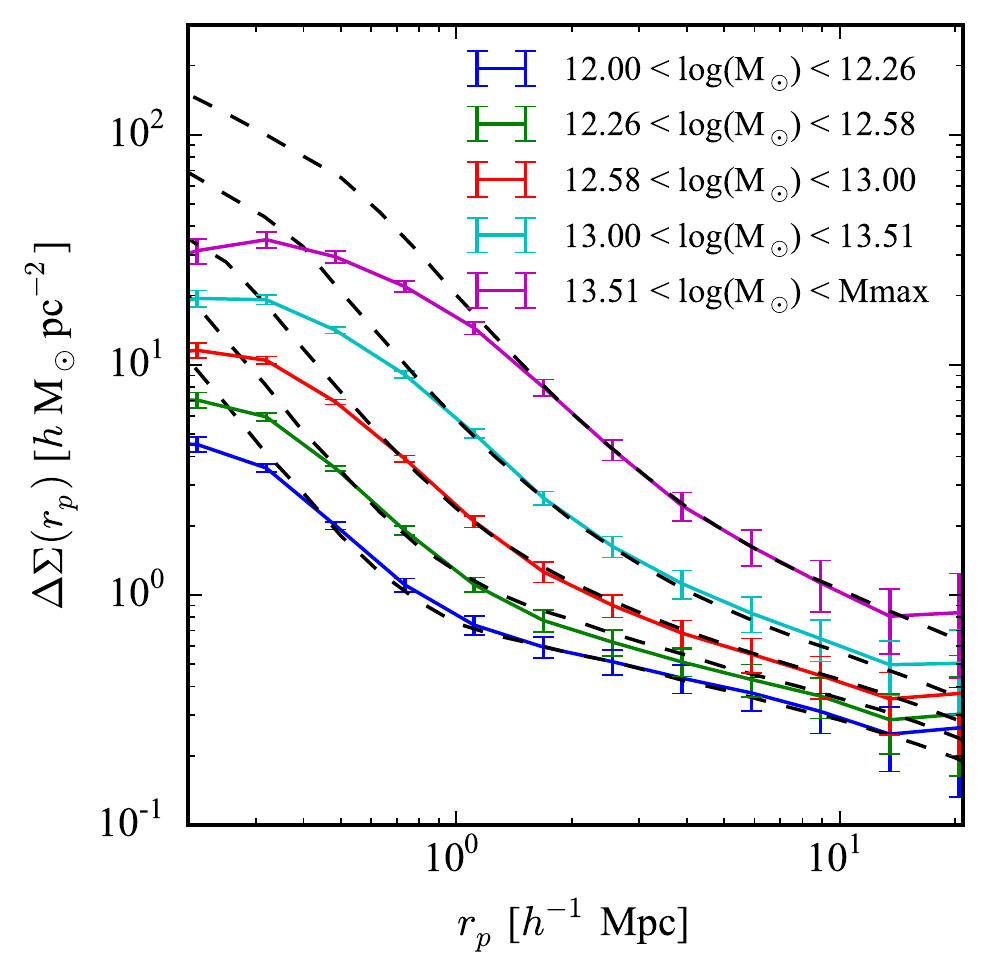}
\end{tabular}
\caption{Same as Figure~\ref{fig:gglensing} but at redshift $z=0.8$. \label{fig:gglensing_highz}}
\end{figure*}

We study  in this  section the  cross-correlation signal  between halo
positions and  the surrounding  density.  For  this, we  identify dark
matter  halos  and remap  their  positions  into cuboidal  coordinates
similarly as previously done with  the particle density in building up
the   light-cones.   We   then    compute   the   halo-shear   lensing
cross-correlation in comoving coordinates as:
\begin{equation}
\Delta \Sigma (r_p) = \gamma_t (r_p) \Sigma_{\rm crit},
\end{equation}
from the shear catalogs of background source galaxies surrounding each
halo  centres  in  projection.  To  account  for  the  different  mass
over-density distributions, we divide halos into five mass bins ranging
from $10^{12}$  to $10^{15}$  $h^{-1} \rm{M_\odot}$.  We  also compute
$\Delta  \Sigma(r_p)$  directly  from  the projected  density  map  by
stacking  the particle  density around  the same  halos. The  averaged
$\Delta  \Sigma(r_p)$ among  the various  light-cone realisations  are
shown           in            Figure~\ref{fig:gglensing}           and
Figure~\ref{fig:gglensing_highz}    for     $z=0.52$    and    $z=0.8$
respectively.  Error  bars  correspond to  standard  deviations  among
light-cone realisations.
These  signals  are compared  with  halo  model predictions.  For  the
latter,   we    used   the   detailed   prescription    presented   in
\citet{vandenbosch13},  which accurately  includes the  halo exclusion
effect  (in computing  the two  halo term  two haloes  cannot be  less
distant than the sum of their  virial radii): distinct haloes. In the model, we assumed
NFW halo radial density profiles with a concentration-mass relation as
consistently computed  by \citet{prada12} and  used the halo  mass and
bias functions by \citet{tinker10}.

First, we  found that our cross-correlation  and stacking measurements
are in  good agreement,  meaning that our  lensing procedure  does not
introduce any significant noise or  smoothing. Second, we found a good
agreement with theory on large  scale, but some discrepancies on small
scale.  The dependence of the discrepancies on the halo mass is due 
to the density contrast estimator $\Delta \Sigma$ applied on the
actual density profile $\Sigma(R)$. The amplitude of the discrepancy
depends on the map resolution but also on the steepness of the density profile
 above the background
(i.e. the 2-halo term). For the high mass halos, the steep NFW
 tail is above the background over a large radial range, whereas for the
 low mass halos, the profile is buried in the background for the most part, and only
 the inner and shallower part can be distinguished from
 the background. At very small scale, where the density signal $\Sigma(R)$ in the
 maps is not properly resolved, the measured density contrast 
 $\Delta \Sigma(R)$ drops.
We performed a test by degrading the map by a factor of 4 and observed
that the discrepancies were shifted to larger radii in agreement with this explanation.

At redshift  $z=0.52$, the difference
between the stacking and the  cross-correlation measurements is due to
the shot  noise in the  maps, which  increases at lower  redshift.  We
also  have  quantified  the   impact  of  shear  versus  reduced-shear
measurements, but at these scales there is no differences even for the
highest  halo-mass  bins.   In   summary,  the  discrepancies  between
theoretical  and measured  curves in  figures \ref{fig:gglensing}  and
\ref{fig:gglensing_highz} at  small angular separation  are consistent
with numerical resolution limitations.

\section{Summary and Conclusion} \label{sumandcon}

In the context of forthcoming large spectroscopic and imaging wide
field surveys that aim at high-precision cosmology, it is mandatory to
have precise numerical simulations in order to test the methods of
analysis, evaluate their predictive power, and estimate errors in the
observables.

In this paper, we present the simulations that we are using in several
forthcoming papers in which we  cross-correlate weak lensing and other
observables.
The     produced     light-cones     are    extracted     from     the
$(2.5\ \mathrm{Gpc}/h)^3$ BigMDPL cosmological simulation.  They
have  been designed  so  that they  match the  shape  of the  publicly
available  VIPERS fields  W1 and  W4, and  cover all  redshifts up  to
$z  = 2.3$.  In total,  we produced  845 deg$^2$,  and 871  deg$^2$ of
independent light-cone realisations for W1 and W4 respectively. All of
them  include  both  lensing  and   halo  mock  catalogs  with  masses
$M_{200} > 10^{12} M\odot/h$.

We have  performed several tests  including cosmic shear  2-points and
3-points statistics, as well as halo-galaxy lensing test for different
bins  of  mass  and  redshifts,  and  we  found  good  agreement  with
theoretical predictions down to the scales numerically resolved by the
BigMDPL  simulation.  In particular, the  converge power spectra
have  also been  compared with  the analytic  predictions from  CAMB
finding  small departures  well  below $5\%$  for  all the  considered
source redshifts, that should be taken into account in future modelling
for the coming area of precision cosmology.

We provide  a first release of  the lensing and halo  mock catalogs on
the       \emph{Bologna       Lens       Factory}       web       site
\footnote{\url{https://bolognalensfactory.wordpress.com/home-2/multdarklens}}.
Tools and  particular statistics  are also  available upon  request as
well as lensing catalogs for specific source redshift distribution.

\section*{Acknowledgements}
The   BigMDPL  simulation   has   been  performed   on  the   SuperMUC
supercomputer at the Leibniz-Rechenzentrum  (LRZ) in Munich, using the
computing resources  awarded to  the PRACE project  number 2012060963.
CG thanks CNES  for financial support.  CG and RBM's  research is part
of the  project GLENCO,  funded under  the European  Seventh Framework
Programme, Ideas, Grant Agreement n.  259349.  GY acknowledges support
from   MINECO  (Spain)   under  research   grants  AYA2012-31101   and
FPA2012-34694 and Consolider Ingenio SyeC CSD2007-0050 CG, EJ and Sdlt
acknowledge the support of the  OCEVU Labex (ANR-11-LABX-0060) and the
A*MIDEX project  (ANR-11-IDEX-0001-02) funded by  the "Investissements
d'Avenir" French government program managed  by the ANR.  We thank the
Red Espa\~nola de  Supercomputaci\' on for granting  us computing time
in  the Marenostrum  Supercomputer at  the BSC-CNS  where part  of the
analyses presented  in this paper  have been performed.  We  thank the
anonymous reviewer for  his/her careful reading of  the manuscript and
comments.  \appendix

\bibliographystyle{mn2e}
\bibliography{paper}
%\bibliography{../../globalbibs.bib}
\label{lastpage}
\end{document}